\documentclass[notitlepage,onecolumn,10pt,nofootinbib]{revtex4-2}
\usepackage[english]{babel}
\usepackage[latin1]{inputenc}
\usepackage{amsmath,amssymb}
\usepackage{bbm}
\usepackage{graphicx}
\usepackage{color}

\begin{document}

\title{On Carrollian and Galilean contractions of BMS algebra in 3 and 4 dimensions}

\author{Andrzej Borowiec}
\affiliation{University of Wroc\l{}aw, Institute of Theoretical Physics, pl.\ M.\ Borna 9, 50-204 Wroc\l{}aw, Poland}
%\emailAdd{andrzej.borowiec@uwr.edu.pl}
\author{Jerzy Kowalski-Glikman}
\affiliation{University of Wroc\l{}aw, Institute of Theoretical Physics, pl.\ M.\ Borna 9, 50-204 Wroc\l{}aw, Poland}
\affiliation{National Centre for Nuclear Research, ul.\ Pasteura 7, 02-093 Warsaw, Poland}
%\emailAdd{jerzy.kowalski-glikman@uwr.edu.pl}
\author{Tomasz Trze\'{s}niewski}
\email{tomasz.trzesniewski@uwr.edu.pl}
\affiliation{University of Wroc\l{}aw, Institute of Theoretical Physics, pl.\ M.\ Borna 9, 50-204 Wroc\l{}aw, Poland}

\date{\today}

\begin{abstract}
In this paper, we find a class of Carrollian and Galilean contractions of (extended) BMS algebra in 3+1 and 2+1 dimensions. To this end, we investigate possible embeddings of 3D/4D Poincar\'{e} into the BMS${}_3$ and BMS${}_4$ algebras, respectively. The contraction limits in the 2+1-dimensional case are then enforced by appropriate contractions of its Poincar\'{e} subalgebras. In 3+1 dimensions, we have to apply instead the analogy between the structures of Poincar\'{e} and BMS algebra. In the case of non-vanishing cosmological constant in 2+1 dimensions, we consider the contractions of $\Lambda$-BMS${}_3$ algebra in an analogous manner.
\end{abstract}

\maketitle

\section{Introduction}

The symmetries of space and time are concepts at the heart of modern physics. Poincar\'{e} symmetry is the cornerstone of classical relativistic physics and quantum field theory. Diffeomorphism invariance, expressed by the equivalence principle is the fundamental concept underlying general relativity. As shown by Arnowitt, Deser and Misner \cite{Arnowitt:1962hi}, and Regge and Teitelboim \cite{Regge:1974zd} in the case of asymptotically flat spacetime, at spatial infinity, the diffeomorphisms quite expectantly reduce to Poincar\'{e} symmetry. It came as a great surprise when shortly afterwards it was discovered that the relevant symmetry algebra at null infinity is not Poincar\'{e} but the infinite-dimensional BMS (Bondi-van der Burg-Metzner-Sachs) algebra \cite{Bondi:1962gm,Sachs:1962gs}. After decades when asymptotic symmetries have been largely forgotten, a few years ago they reemerged as one of the most actively studied topics in high energy physics and gravity. The breakthrough here was the research carried out by Strominger and his collaborators that resulted in revealing a close relation between BMS symmetry, Weinberg's soft theorems, and memory effect (see \cite{Strominger:2017zoo} for a review). 

The original BMS algebra is a semidirect product of the standard Lorentz algebra and the infinite-dimensional commuting algebra of supertranslations. This algebra contains the Poincar\'{e} subalgebra, spanned by the Lorentz algebra and four supertranslation generators. It should be noted however that the form of asymptotic symmetry algebra depends crucially on the adopted asymptotic conditions for the spacetime metric. Indeed, it was shown in \cite{Barnich:2010sd,Barnich:2010eb} that by relaxing these boundary conditions, one can extend BMS algebra so as to make it composed of infinite supertranslational and superrotation sectors. From a different point of view, the original BMS algebra becomes extended by superrotations when we go down the tower of subleading asymptotic structures \cite{Strominger:2017zoo,Geiller:2024cs}. Meanwhile, it turned out that by relaxing boundary conditions at spatial infinity, the asymptotic symmetry algebra there can also be extended from Poincar\'{e} to BMS \cite{Henneaux:2018cst,Brocki:2021ieh}. Similarly, the asymptotic symmetries can be made to emerge at non-extremal horizons \cite{Grumiller:2020sr}. 

The 2+1 dimensional counterpart of BMS algebra in 3+1 dimensions has also attracted a lot of attention. It was originally analyzed in \cite{Brown:1986nw} and these investigations are recognized as a predecessor of the AdS/CFT research program. Comparing the derivations for the 2+1- and 3+1-dimensional cases is very instructive, see \cite{Barnich:2007cs}. Recently, the BMS${}_3$ was investigated in detail \cite{Barnich:2014kra,Barnich:2015uva}. From our perspective, the case of BMS${}_3$ is particularly interesting because this algebra can be extended to the case of non-vanishing cosmological constant \cite{Parsa:2019os}, which is impossible in the case of BMS${}_4$ \cite{Safari:2019os}, where the corresponding structure becomes a Lie algebroid \cite{Compere:2019ts}. 

In all the cases discussed above, Poincar\'{e} algebra played a crucial role, being the largest finite-dimensional subalgebra of BMS algebra. It is well known however that Poincar\'{e} algebra is not a unique algebra to describe the kinematics of particles and fields. In fact, the classification of such 'kinematical algebras' (see \cite{Figueroa:2022ns,Bergshoeff:2023ar} for reviews) was provided in the seminal work by Bacry and L\'{e}vy-Leblond \cite{Bacry:1968ps}, later extended in \cite{Bacry:1986cy} and generalized to 2+1 dimensions only recently in \cite{Andrzejewski:2018ks}. The idea was to find a set of 10-parameter algebras that contains space and time translations, rotations and boosts, satisfying some physically appealing conditions: the isotropy of space (the standard action of rotations on energy, momenta, and boosts), the invariance under discrete symmetries of time reversal and parity (this condition can be lifted, as done in \cite{Bacry:1986cy}), and the noncompactness of boosts -- the transformations between inertial observers. Among the kinematical algebras, apart from the Lorentzian algebras (i.e., Poincar\'{e} or (Anti)-de Sitter), two families are particularly physically relevant: the Galilean and Carrollian one \cite{Duval:2014ce,Figueroa:2023ly}. Both can be obtained as contraction limits of Lorentzian algebras, with speed of light being the contraction parameter; the limits are opposite, however: $c \rightarrow \infty$ in the Galilean case and $c \rightarrow 0$ in the Carrolian one. 

The investigation of these regimes in the theory of gravity, where they are closely related to expansions around the weak and strong gravitational coupling, respectively, is currently a subject of great interest, see e.g. \cite{Bergshoeff:2017cy,Hansen:2020nr,Hansen:2022cy}, or a review \cite{Hartong:2023ry} and references therein (in particular, it was finally found \cite{Bleeken:2017ty,Hansen:2019ay,Hansen:2020nr} how to correctly recover Newtonian gravity). Similar studies have been developed in 2+1-dimensional gravity \cite{Papageorgiou:2009as,Papageorgiou:2010ga,Matulich:2019ln,Gomis:2020ny}. On the other hand, a geometry with the Galilean symmetries can also be obtained via the null reduction of a Lorentzian geometry \cite{Duval:1985by,Gibbons:2003nt}, while null hypersurfaces in a Lorentzian geometry are geometries with the Carrollian symmetries \cite{Duval:2014ce,Figueroa:2019ss}. Moreover, such structures have applications going beyond general relativity, including non-AdS holography \cite{Donnay:2023by,Taylor:2016ly}. 

Since Poincar\'{e} algebra is contained in BMS algebra, it should also be possible to extend its Carrollian and Galilean contractions to the latter. This has recently been achieved \cite{Perez:2021ay,Fuentealba:2022as} for the original BMS, while the aim of the present paper is to define such contractions for the extended BMS algebra, both in 3+1 and 2+1 spacetime dimensions. 

The Galilean contraction of BMS algebra corresponds physically to the non-relativistic limit of the theory. This contraction does not seem to be of any physical relevance in the case of the symmetries of asymptotically flat gravitational systems in the neighborhood of null infinity in four spacetime dimensions. It is however of physical interest in the case of 3+1-dimensional asymptotically flat Newtonian systems at spatial infinity. Contrary to that, the Carrollian contraction is of physical interest not only in the neighbourhood of null infinity but also in describing gravity on manifolds with null boundaries (see e.g., \cite{Perez:2021ay}, \cite{Fuentealba:2022as}, \cite{Freidel:2023bnj}, \cite{Ciambelli:2023mir}, \cite{Adami:2023ce} and references therein). This should not be confused with an interesting fact that the original BMS algebra itself turns out to be a conformal extension of the lower-dimensional Carroll algebra \cite{Duval:2014cs,Ciambelli:2019cs}. 

The plan of this paper is as follows. In the following Section \ref{sec:21}, we shortly recall the Carrollian and Galilean contractions of (finite-dimensional) kinematical algebras for any value of the cosmological constant, taking 2+1 spacetime dimensions as an example. Section \ref{sec:3.0} is devoted to a discussion of BMS${}_3$ algebra and the embeddings of the (2+1)d Poincar\'{e} algebra into it. Then, in Section \ref{sec:4.0}, we discuss contractions of 2+1-dimensional BMS algebras without and (Subsection \ref{sec:4.1}) with the cosmological constant, while in Section \ref{sec:5.0}, we consider the 3+1 dimensional (extended) BMS algebra and its contraction limits. Finally, Section \ref{sec:6.0} summarizes the obtained results.

\section{Carrollian and Galilean kinematical Lie algebras in (2+1)d}\label{sec:21}

The Lie brackets of 2+1-dimensional Poincar\'{e} and (Anti-)de Sitter algebras, $\mathfrak{iso}(2,1) = \mathfrak{so}(2,1) \vartriangleright\!\!< {\cal T}^{2,1}$, $\mathfrak{so}(3,1)$ and $\mathfrak{so}(2,2)$ (for the cosmological constant $\Lambda = 0$, $\Lambda > 0$ or $\Lambda < 0$, respectively), can be expressed in a unified fashion:
\begin{align}\label{eq:32.03a}
[{\cal J}_\mu,{\cal J}_\nu] = \epsilon_{\mu\nu}^{\ \ \sigma} {\cal J}_\sigma\,, \qquad [{\cal J}_\mu,{\cal P}_\nu] = \epsilon_{\mu\nu}^{\ \ \sigma} {\cal P}_\sigma\,, \qquad [{\cal P}_\mu,{\cal P}_\nu] = -\Lambda\, \epsilon_{\mu\nu}^{\ \ \sigma} {\cal J}_\sigma\,,
\end{align}
where $\mu = 0,1,2$, $\epsilon_{012} = 1$ and indices are raised with Minkowski metric $(1,-1,-1)$). Let us perform a change of basis
\begin{align}\label{eq:10x}
J := -{\cal J}_0\,, \qquad K_a := -{\cal J}_a\,, \qquad P_0 := {\cal P}_0\,, \qquad P_{1/2} := \mp {\cal P}_{2/1}\,,
\end{align}
so that the brackets \eqref{eq:32.03a} become
\begin{align}\label{eq:10}
[J,K_a] & = \epsilon_a^{\ b} K_b\,, & [K_1,K_2] & = -J\,, & 
[J,P_a] & = \epsilon_a^{\ b} P_b\,, & [J,P_0] & = 0\,, \nonumber\\ 
[K_a,P_b] & = \delta_{ab} P_0\,, & [K_a,P_0] & = P_a\,, & [P_1,P_2] & = \Lambda\, J\,, & [P_0,P_a] & = -\Lambda\, K_a\,,
\end{align}
where $a = 1,2$ and $\epsilon_1^{\ 2} = 1$. In order to perform the Carrollian contraction of either of the considered Lorentzian algebras, one needs to denote $R := J$, ${\cal T}_a := P_a$, define the rescaled generators $Q_a := c\, K_a$, ${\cal T}_0 := c\, P_0$ and take the limit $c \to 0$. As a result, we obtain the brackets of 2+1-dimensional Carroll (for $\Lambda = 0$) or (Anti-)de Sitter-Carroll algebra (also called the para-Poincar\'{e} if $\Lambda < 0$ and para-Euclidean if $\Lambda > 0$, due to their respective isomorphisms with 2+1-dimensional Poincar\'{e} and Euclidean algebras, see e.g. \cite{Trzesniewski:2023qi}):
\begin{align}\label{eq:11d}
[R,Q_a] & = \epsilon_a^{\ b} Q_b\,, & [Q_1,Q_2] & = 0\,, & 
[R,{\cal T}_a] & = \epsilon_a^{\ b} {\cal T}_b\,, & [R,{\cal T}_0] & = 0\,, \nonumber\\ 
[Q_a,{\cal T}_b] & = \delta_{ab} {\cal T}_0\,, & [Q_a,{\cal T}_0] & = 0\,, & [{\cal T}_1,{\cal T}_2] & = \Lambda\, R\,, & [{\cal T}_0,{\cal T}_a] & = -\Lambda\, Q_a\,.
\end{align}
On the other hand, the Galilean contraction of a Lorentzian algebra consists in denoting $R := J$, ${\cal T}_0 := P_0$, introducing the rescaled generators $Q_a := c^{-1} K_a$, ${\cal T}_a := c^{-1} P_a$ and taking the limit $c \to \infty$. It allows to obtain the brackets of 2+1-dimensional Galilei (for $\Lambda = 0$) or (Anti-)de Sitter-Galilei algebra (also called the oscillating Newton-Hooke if $\Lambda < 0$ and the expanding Newton-Hooke if $\Lambda > 0$):
\begin{align}\label{eq:11e}
[R,Q_a] & = \epsilon_a^{\ b} Q_b\,, & [Q_1,Q_2] & = 0\,, & 
[R,{\cal T}_a] & = \epsilon_a^{\ b} {\cal T}_b\,, & [R,{\cal T}_0] & = 0\,, \nonumber\\ 
[Q_a,{\cal T}_b] & = 0\,, & [Q_a,{\cal T}_0] & = {\cal T}_a\,, & [{\cal T}_1,{\cal T}_2] & = 0\,, & [{\cal T}_0,{\cal T}_a] & = -\Lambda\, Q_a\,.
\end{align}
Finally, let us recall that Carroll and Galilei algebras are also directly related with Poincar\'{e} algebra one dimension higher \cite{Figueroa:2023ly}. In particular, in the 2+1-dimensional case, Carroll algebra can be embedded as a subalgebra of (3+1)d Poincar\'{e} (see \cite{Trzesniewski:2023qi} for the explicit formulae), while Galilei algebra can be obtained as a quotient subalgebra of (3+1)d Poincar\'{e}.

\section{BMS${}_3$ algebra $\mathfrak{B}_3$, its real forms and embeddings of $\mathfrak{iso}(2,1)$}\label{sec:3.0}

The brackets of BMS algebra in 2+1 dimensions (which we will denote by $\mathfrak{B}_3$) in terms of the generators of superrotations $l_{n}$ and supertranslations $T_{n}$ have the form
\begin{align}\label{BMS3}
\left[l_{n}, l_{m}\right] = (n - m)\, l_{n + m}\,, \qquad \left[l_{n}, T_{m}\right] = (n - m)\, T_{n + m}\,,
\end{align}
where $n, m \in \mathbb{Z}$, while the supertranslation subalgebra is commutative. 

From the mathematical perspective, $\mathfrak{B}_3$ is also known as the inhomogeneous centreless two-sided Witt algebra, $\mathfrak{W}(2,2)$. Moreover, it is actually a complex algebra and in order to impose the reality conditions on its generators, one needs to introduce a real structure, i.e. a $*$-conjugation (an involutive, antilinear antiautomorphism). The $\mathfrak{B}_3$ algebra has four such possible real forms, inherited from real forms of the $\mathfrak{sl}(2,\mathbb{C})$ algebra:
%\footnote{All non-compact real forms of $\mathfrak{sl}(2,\mathbb{C})$ ($\mathfrak{sl}(2,\mathbb{R})$, $\mathfrak{su}(1,1)$ and $\mathfrak{so}(2,1)$) are known to be isomorphic, while it is no longer true for the real forms of $\mathfrak{B}_3$.}
\begin{itemize}
    \item[i)] type $\mathfrak{sl}(2,\mathbb{R})$, with the reality conditions $\ l_n^* = -l_n\,, \quad T_n^* = -T_n\,, \quad n \in \mathbb{Z}\,$,
    \item[ii)] type $\mathfrak{su}(1,1)$, with the reality conditions $\ l_n^* = l_{-n}\,, \quad T_n^* = T_{-n}\,, \quad n \in \mathbb{Z}\,$,
    \item[iii)] type $\mathfrak{so}(2,1)$, with the reality conditions $\ l_n^* = (-1)^{n+1} l_n\,, \quad T_n^* = (-1)^{n+1} T_n\,, \quad n \in \mathbb{Z}\,$,
    \item[iv)] type $\mathfrak{su}(1,1)|\mathfrak{su}(2)$, with the reality conditions $\ l_n^* = (-1)^n l_{-n}\,, \quad T_n^* = (-1)^n T_{-n}\,, \quad n \in \mathbb{Z}\,$,
\end{itemize}
Similarly to the corresponding (non-compact) real forms of $\mathfrak{sl}(2,\mathbb{C})$, it can be shown that the cases i-iii) are isomorphic to each other. The reality conditions of the $\mathfrak{su}(2)$-type ($\mathfrak{su}(2)$ is the remaining, compact real form of $\mathfrak{sl}(2,\mathbb{C})$): $l_0^* = l_0\,, \ l_n^* = -l_{-n}\,, \ n \neq 0\,$, are incompatible with the first bracket \eqref{BMS3} for $n \pm m \neq 0$ and hence they are also not valid for the full $\mathfrak{B}_3$ algebra. However, case iv) combines the real structure of type $\mathfrak{su}(1,1)$ for even $n$ and type $\mathfrak{su}(2)$ for odd $n$. \medskip\\
\indent Let us now recall that the non-vanishing brackets of Poincar\'{e} algebra in arbitrary dimension $d$+1 (with the flat metric $\eta$ of any signature and generators satisfying the anti-Hermitian reality conditions $X^* = -X$) can be written as
\begin{align}\label{Poincare}
\left[M_{\mu\nu}, M_{\rho\sigma}\right] &= \eta_{\mu\sigma} M_{\nu\rho} + \eta_{\nu\rho} M_{\mu\sigma} - \eta_{\mu\rho} M_{\nu\sigma} - \eta_{\nu\sigma} M_{\mu\rho}\,, \nonumber\\
\left[M_{\mu\nu}, P_\rho\right] &= \eta_{\nu\rho} P_\mu - \eta_{\mu\rho} P_\nu\,,
\end{align}
where indices run from 0 to $d$. In particular, we may change the basis in 2+1 dimensions to $J = M_{12}$, $K_{i} = M_{i0}$, $P_{0}$, $P_{i}$, with the diagonal metric $(\eta_{00}, \eta_{11}, \eta_{22})$, so that \eqref{Poincare} becomes
\begin{alignat}{4}\label{eq:22.00}
\left[J, K_{i}\right] &= -\eta_{ii} \epsilon_{ij} K_{j}\,, &\qquad \left[K_i, K_{j}\right] &= -\eta_{00} \epsilon_{ij} J\,, && \nonumber\\
\left[J, P_{i}\right] &= -\eta_{ii} \epsilon_{ij} P_{j}\,, &\qquad
\left[K_{i}, P_{j}\right] &= -\eta_{ij} P_{0}\,, &\qquad \left[K_{i}, P_{0}\right] &= \eta_{00} P_{i}\,,
\end{alignat}
which is a generalization of \eqref{eq:10} for $\Lambda = 0$. \medskip\\
\indent There exists a one-parameter family ${\cal P}_{n}(1,2) = {\rm span}\{l_{0}, l_{\pm n}, T_{0}, T_{\pm n}\}$, $n \in \mathbb{N}$ (in this paper we adopt the convention $0 \neq \mathbb{N}$) of maximal finite-dimensional subalgebras of $\mathfrak{B}_3$, each of them isomorphic to the Poincar\'{e} algebra $\mathfrak{iso}(2,1)$. In particular, if we equip $\mathfrak{B}_3$ with real structure of the $\mathfrak{su}(1,1)$-type, such a Poincar\'{e} subalgebra can be described as the image of an embedding ${\rm P}^{(n)}(1,2)$ of $\mathfrak{iso}(2,1)$ into $\mathfrak{B}_3$, spanned by the generators:
%Equivalently, we can define a family of embeddings ${\rm P}^{(n)}(1,2) = \{J, K_{i(n)}, P_0, P_{i(n)}\,;\ i = 1,2\}$ of $\mathfrak{iso}(2,1)$ into $\mathfrak{B}_3$, given by:
\begin{alignat}{4}\label{eq:22.01}
J &= i\, l_{0}\,, &\qquad K_1^{(n)} &= \frac{1}{2} \left(l_{n} - l_{-n}\right), &\qquad K_2^{(n)} &= -\frac{i}{2}\left(l_{n} + l_{-n}\right), \nonumber\\
P_{0} &= i\, T_{0}\,, &\qquad P_1^{(n)} &= \frac{i}{2} \left(T_{n} + T_{-n}\right), &\qquad P_2^{(n)} &= \frac{1}{2}\left(T_{n} - T_{-n}\right),
\end{alignat}
which satisfy the brackets \eqref{eq:22.00} with the metric $\eta = n\, {\rm diag}(1,-1,-1)$ (if needed, the factor $n$ can eliminated with the help of rescaling of $J$, $K_{i(n)}$ by $1/n$ and $P_0$, $P_{i(n)}$ by $n$) and the anti-Hermitian reality conditions, i.e. $X^* = -X$. The family of all embeddings ${\rm P}^{(n)}(1,2)$ determines a special basis of the $\mathfrak{B}_3$ algebra $\bigcup_{n \in \mathbb{N}} \{K_1^{(n)}, K_2^{(n)}, P_1^{(n)}, P_2^{(n)}\} \cup \{J, P_0\}$, which will play the central role in defining the Carrollian/Galilean contractions of the latter. 
%Such embeddings are consistent with equipping $\mathfrak{B}_3$ with reality conditions of the $\mathfrak{su}(1,1)$-type, i.e. the latter lead to the anti-Hermitian reality conditions ($X^* = -X$) for each set of Poincar\'{e} generators \eqref{eq:22.01}. Moreover, the full family corresponds to a particular basis $\bigcup_{n \in \mathbb{N}} \{K_1^{(n)}, K_2^{(n)}, P_1^{(n)}, P_2^{(n)}\} \cup \{J, P_0\}$ of the $\mathfrak{B}_3$ algebra. 

If we consider real form of the $\mathfrak{sl}(2,\mathbb{R})$-type instead, Poincar\'{e} subalgebras can be identified with another family of embeddings ${\rm P}'^{(n)}(1,2)$, spanned by the anti-Hermitian generators:
%On the other hand, if we choose reality conditions of the $\mathfrak{sl}(2,\mathbb{R})$-type, Poincar\'{e} generators with the anti-Hermitian reality conditions are obtained for another family of embeddings, ${\rm P}'^{(n)}(1,2) = \{J^{(n)}, K_1^{(n)}, P_0^{(n)}, P_1^{(n)}, K_2, P_1\}$ defined as:
\begin{alignat}{4}\label{eq:22.02}
J^{(n)} &= \frac{1}{2} \left(l_{n} + l_{-n}\right), &\qquad K_1^{(n)} &= \frac{1}{2} \left(l_{n} - l_{-n}\right), &\qquad K_2 &= -l_0\,, \nonumber\\
P_0^{(n)} &= \frac{1}{2} \left(T_{n} + T_{-n}\right), &\qquad P_2^{(n)} &= \frac{1}{2} \left(T_{n} - T_{-n}\right), &\qquad P_1 &= T_0\,,
\end{alignat}
which satisfy the brackets \eqref{eq:22.00} with the metric $\eta = n\, {\rm diag}(1,-1,-1)$. The family corresponds to a differently arranged basis of the $\mathfrak{B}_3$ algebra $\bigcup_{n \in \mathbb{N}} \{J^{(n)}, K_1^{(n)}, P_0^{(n)}, P_2^{(n)}\} \cup \{K_2, P_1\}$. Consequently, as we will see, the contractions performed using this basis work in a different way than for the $\mathfrak{su}(1,1)$-type real form. 

%Summarizing, for each $n \in \mathbb{N}$, there exist at least two inequivalent embeddings of (real) Poincar\'{e} algebra $\mathfrak{iso}(2,1)$ into $\mathfrak{B}_3$. 
Both families of embeddings allow us to decompose (a real form of) $\mathfrak{B}_3$ into the union of its maximal finite-dimensional subalgebras isomorphic to $\mathfrak{iso}(2,1)$, which have the non-empty intersection spanned by $l_0$ and $T_0$. Real forms of the types $\mathfrak{so}(2,1)$ and $\mathfrak{su}(1,1)|\mathfrak{su}(2)$ are less appealing in the sense that they are inconsistent with such a decomposition into Poincar\'{e} subalgebras covering the whole $\mathfrak{B}_3$. A family of embeddings associated to the former real form is given by ${\rm P}'^{(n)}(1,2)$ but with $n \in 2\mathbbm{N}$, while embeddings associated to the latter are ${\rm P}^{(n)}(1,2)$ with $n \in 2\mathbbm{N}$. %Let us also note that $\mathfrak{iso}(3)$ cannot be embedded into $\mathfrak{B}_3$ because there is no real form of the $\mathfrak{su}(2)$-type. 
\medskip\\
Let us note in passing that one can construct vacuum spacetimes associated with different choices of an embedding of $\mathfrak{iso}(2,1)$ into $\mathfrak{B}_3$. To this end, we consider the general solution of the vacuum Einstein equations in the Bondi gauge \cite{Barnich:2010eb}
\begin{align}
ds^2 = \Theta(\phi)\, du^2 - 2 du dr + \left(u\, \Theta'(\phi) + \Xi(\phi)\right) du d\phi + r^2 d\phi^2\,,
\end{align}
where $\Theta$, $\Xi$ are arbitrary periodic functions. It is then convenient to introduce a new variable $z = e^{i \phi}$, bringing the metric to the form
\begin{align}\label{metric3dbondi}
ds^2 = \Theta(z)\, du^2 - 2du dr + \left(u\, \Theta'(z) - i\, \Xi(z) z^{-1}\right) du dz - \frac{r^2}{z^2}\, dz^2\,.
\end{align}
Now, we want to find the Poincar\'{e} vacua, i.e. spacetimes whose metric, of the general form \eqref{metric3dbondi}, has the maximal set of six Killing vector fields corresponding to one of the embeddings \eqref{eq:22.01} or \eqref{eq:22.02}. This problem was analyzed in \cite{Borowiec:2021bs}, where it was shown that the conditions for the vector fields corresponding to the generators $l_n$, $T_n$ to be Killing vectors of the metric \eqref{metric3dbondi} are
\begin{align}\label{Killingcond}
\Xi = 0\,, \quad \Theta = -n^2\,.
\end{align}
We see that since the embeddings \eqref{eq:22.01} or \eqref{eq:22.02} differ only be arrangement of the $l_n$ and $T_n$ generators, they both lead to the same family of vacuum spacetimes. In other words, there is a degeneracy in the mapping of vacuum spacetimes to embeddings of the (local) Poincar\'{e} symmetry into the asymptotic BMS symmetry. 

It is also worthy to note that the vacuum spacetimes satisfying \eqref{Killingcond} generically (apart from $n = \pm 1$) possess a conical singularity \cite{Borowiec:2021bs} (see also \cite{Barnich:2015uva}), i.e. their metric can be expressed as
\begin{align}
ds^2 = -dt^2 + dr^2 + r^2 n^2 d\phi^2\,.
\end{align}
Thus, if $n \neq \pm 1$, we have here to do with (2+1)d spacetimes containing a massive particle (cf. \cite{Matschull:1997du}) with the quantized, negative mass $M = -(|n|-1 )/(4G)$, where $G$ is the three-dimensional Newton's constant. The physical meaning of such vacua is not completely clear and deserves further studies.

\section{Contractions of the BMS${}_3$ and $\Lambda$-BMS${}_3$ algebras}\label{sec:4.0}

%\subsection{BMS${}_3$}

Based on the family of embeddings \eqref{eq:22.01} or \eqref{eq:22.02}, associated with the respective real form, we want to extend the Carrollian and Galilean contractions of Poincar\'{e} algebra $\mathfrak{iso}(2,1)$ discussed in Sec.~\ref{sec:21} to the BMS${}_3$ algebra $\mathfrak{B}_3$ \eqref{BMS3}. Our guiding principle is that each subalgebra of the contracted $\mathfrak{B}_3$ that was isomorphic to $\mathfrak{iso}(2,1)$ before a contraction (within the chosen family of embeddings) should be isomorphic to Carroll \eqref{eq:11d} or Galilei \eqref{eq:11e} algebra after the respective contraction. An equivalent point of view is to consider this a consistent extension of Carroll/Galilei algebra by superrotations and supertranslations, which aligns with the approach of \cite{Fuentealba:2022as} to the Carrollian and Galilean contractions of the original (non-extended) BMS${}_4$ algebra. In practice, it means that, in the contraction procedure, we need to appropriately rescale all generators of $\mathfrak{B}_3$ that play the role of generators of boosts and time translation (in the Carrollian case) or generators of boosts and spatial translations (in the Galilean case) in any subalgebra spanned by \eqref{eq:22.01} / \eqref{eq:22.02}. \medskip\\ 
Let us first consider the family \eqref{eq:22.01}, which corresponds to the choice of the $\mathfrak{su}(1,1)$-type real form. It can be shown that the rescaling of generators of Poincar\'{e} subalgebras can equivalently be applied to the standard basis of $\mathfrak{B}_3$, i.e. $\{l_n,T_n;\, n \in \mathbb{Z}\}$ with the brackets \eqref{BMS3}, and it follows that:
\begin{itemize}
\item in the case of Carrollian contraction, we should perform the rescaling $l_{n} \mapsto c\, l_{n} = \tilde{l}_{n}\,$, $T_{0} \mapsto c\, T_{0} = \widetilde{T}_{0}$ for all $ n \neq 0$ and take the limit $c \to 0$ to obtain
\begin{alignat}{4}\label{BMS3C}
\big[\tilde{l}_{n}, \tilde{l}_{m}\big] &= 0\,, &\qquad \big[l_{0}, \tilde{l}_{n}\big] &= -n\, \tilde{l}_{n}\,, &\qquad \big[l_{0}, T_{n}\big] &= -n\, T_{n}\,, \nonumber\\
\big[l_{0}, \widetilde{T}_{0}\big] &= 0\,, &\qquad \big[\tilde{l}_{n}, \widetilde{T}_{0}\big] &= 0\,, &\qquad \big[\tilde{l}_{n}, T_{m}\big] &= 2\delta_{m,-n} n\, \tilde{T}_{0}\,,
\end{alignat}
which we will call the Carroll-BMS${}_3$ algebra, CBMS${}_3$; 
\item in the case of Galilean contraction, we perform the rescaling $l_{n} \mapsto c^{-1} l_{n} = \hat{l}_{n}\,$, $T_{n} \mapsto c^{-1} T_{n} = \widehat{T}_{n}$ for all $ n \neq 0$ and take the limit $c \to \infty$ to obtain
\begin{alignat}{4}\label{BMS3G}
\big[\hat{l}_{n}, \hat{l}_{m}\big] &= 0\,, &\qquad \big[l_{0}, \hat{l}_{n}\big] &= -n\, \hat{l}_{n}\,, &\qquad \big[l_{0}, \widehat{T}_{n}\big] &= -n\, \widehat{T}_{n}\,, \nonumber\\
\big[l_{0}, T_{0}\big] &= 0\,, &\qquad \big[\hat{l}_{n}, T_{0}\big] &= n\, \widehat{T}_{n}\,, &\qquad \big[\hat{l}_{n}, \widehat{T}_{m}\big] &= 0\,,
\end{alignat}
which we will call the Galilei-BMS${}_3$ algebra, GBMS${}_3$. 
\end{itemize}
%Comparing the brackets \eqref{BMS3C}, \eqref{BMS3G} with those of Carroll and Galilei algebras (given by \eqref{eq:11d}, \eqref{eq:11e} with $\Lambda = 0$), we observe that indeed, the structures of the considered infinite- and finite-dimensional algebras are completely analogous. In particular, the generators $l_0$, $l_n$, $T_0$ and $T_n$ appear in the same places as the Carroll/Galilei generators of rotation, boosts, time translation and spatial translations, respectively. It is also easy to check that the embeddings of Carroll and Galilei algebras into CBMS${}_3$ and GBMS${}_3$, respectively, can be defined analogously to the embeddings of Poincar\'{e} into BMS${}_3$. 
Comparing the brackets \eqref{BMS3C}, \eqref{BMS3G} with those of (2+1)d Carroll and Galilei algebras (given by \eqref{eq:11d}, \eqref{eq:11e} with $\Lambda = 0$), we observe that indeed, the structures of the considered infinite- and finite-dimensional algebras are completely analogous. In particular, the generators $l_0$, $l_n$, $T_0$ and $T_n$ appear in the same places as the Carroll/Galilei generators of rotation, boosts, time translation and spatial translations, respectively. The underlying reason is that if we change the basis of CBMS${}_3$ / GBMS${}_3$ to the one analogous to \eqref{eq:22.01} or, equivalently, perform either of the contractions in the basis \eqref{eq:22.01}, it turns out that the embeddings of Carroll and Galilei algebras into CBMS${}_3$ and GBMS${}_3$, respectively, can be defined analogously to the embeddings of Poincar\'{e} into the $\mathfrak{B}_3$ algebra. Moreover, in contrast to what happens for $\mathfrak{B}_3$ (the relevant formulae can be obtained by changing the basis in \eqref{eq:22.03} below), all commutation relations between generators of different Carroll/Galilei subalgebras vanish, except the ones involving the two generators that are shared by all of them, $l_0$ and $T_0$ (or $\widetilde{T}_{0}$). If the latter was not the case, CBMS${}_3$ and GBMS${}_3$ would have structure of the direct product of infinite number of Carroll or Galilei subalgebras, respectively. %As a result, CBMS${}_3$ has the structure of $\bigoplus_{n \in \mathbb{N}} \mathfrak{c}_3^{(n)}/\sim$, where each subalgebra $\mathfrak{c}_3^{(n)}$ is isomorphic to Carroll algebra, while GBMS${}_3$ has the structure of $\bigoplus_{n \in \mathbb{N}} \mathfrak{g}_3^{(n)}/\sim$, where each subalgebra $\mathfrak{g}_3^{(n)}$ is isomorphic to Galilei algebra; $\sim$ in both cases denotes an equivalence relation identifying the two generators shared by all subalgebras. 
\medskip\\
\indent If we choose the $\mathfrak{sl}(2,\mathbb{R})$-type real form, the contractions based on the family of embeddings \eqref{eq:22.02} cannot be performed in the basis $\{l_n,T_n;\, n \in \mathbb{Z}\}$ (since e.g. both rotation and boost generators now depend on all $l_n$, $n \in \mathbb{Z}^*$). Instead, we stay in the basis $\{K_2, P_1, K_1^{(n)}, J^{(n)}, P_2^{(n)}, P_0^{(n)};\, n \in \mathbb{N}\}$ and calculate the contraction limits of commutation relations between different Poincar\'{e} subalgebras
\begin{align}\label{eq:22.03}
\left[J^{(n)}, J^{(m)}\right] &= \frac{1}{2} \left((n - m)\, K_1^{(n + m)} + (n + m)\, \sigma_2 K_1^{(|n - m|)}\right)\,, \nonumber\\ \left[J^{(n)}, K_1^{(m)}\right] &= \frac{1}{2} \left((n - m)\, J^{(n + m)} - (n + m)\, J^{(|n - m|)}\right)\,, \nonumber\\
\left[K_1^{(n)}, K_1^{(m)}\right] &= \frac{1}{2} \left((n - m)\, K_1^{(n + m)} - (n + m)\, \sigma_2 K_1^{(|n - m|)}\right)\,, \nonumber\\
\left[J^{(n)}, P_{0/2}^{(m)}\right] &= \frac{1}{2} \left((n - m)\, P_{2/0}^{(n + m)} \pm (n + m)\, \sigma_{2/0} P_{2/0}^{(|n - m|)}\right)\,, \nonumber\\
\left[K_1^{(n)}, P_{0/2}^{(m)}\right] &= \frac{1}{2} \left((n - m)\, P_{0/2}^{(n + m)} \pm (n + m)\, \sigma_{0/2} P_{0/2}^{(|n - m|)}\right),
\end{align}
where $n \neq m$ and, for brevity, $\sigma_0 := 1$, $\sigma_2 := {\rm sgn}(n - m)$ (the remaining brackets for each $n$ are the ones of Poincar\'{e} algebra \eqref{eq:22.00} with $\eta = n\, {\rm diag}(1,-1,-1)$). It turns out that: 
\begin{itemize}
    \item in the case of Carrollian contraction, the rescaling $K_2 \mapsto c\, K_2\,$, $K_1^{(n)} \mapsto c\, K_1^{(n)}\,$, $P_0^{(n)} \mapsto c\, P_0^{(n)}$ (for all $n \in \mathbbm{N}$) and taking the limit $c \to 0$ lead to the divergence of the first and fourth bracket in \eqref{eq:22.03}; 
    \item in the case of Galilean contraction, the rescaling $P_1 \mapsto c^{-1} P_1\,$, $P_2^{(n)} \mapsto c^{-1} P_2^{(n)}\,$, $K_2 \mapsto c^{-1} K_2\,$, $K_1^{(n)} \mapsto c^{-1} K_1^{(n)}$ (for all $n \in \mathbbm{N}$) and taking the limit $c \to \infty$ lead to the divergence of the first and fourth bracket in \eqref{eq:22.03}.
\end{itemize}

Therefore, both contractions are ill-defined. On the other hand, it is still possible to arrive at the algebra \eqref{BMS3C} or \eqref{BMS3G} starting from \eqref{eq:22.02} if we perform a different kind of contractions, which are equivalent to the ones performed for the $\mathfrak{su}(1,1)$-type real form when expressed in the basis $\{l_n,T_n;\, n \in \mathbb{Z}\}$ but their interpretation for the $\mathfrak{sl}(2,\mathbb{R})$-type real form is different. To this end, let us first change the basis of $\mathfrak{B}_3$ from \eqref{eq:22.02} (keeping $P_1$) to
\begin{align}
M_{\pm 1}^{(n)} = \mp \frac{1}{\sqrt{2}} \left(J^{(n)} \pm K_1^{(n)}\right)\,, \qquad M_{+-} = K_2\,, \qquad P_\pm^{(n)} = \frac{1}{\sqrt{2}} \left(P_0^{(n)} \pm P_2^{(n)}\right),
\end{align}
where $n \in \mathbb{N}$. In a given embedding ${\rm P}'^{(n)}(1,2)$ of Poincar\'{e} algebra $\mathfrak{iso}(2,1)$, $M_{\pm 1}^{(n)}$ are actually two generators of null rotations (parabolic Lorentz transformations) and $P_\pm^{(n)}$ are two generators of translations along the null directions. The non-vanishing brackets of (the image of) ${\rm P}'^{(n)}(1,2)$ become
\begin{align}\label{eq:4.1}
[M_{+-}, M_{\pm 1}^{(n)}] &= \pm\eta_{+-} M_{\pm 1}^{(n)}\,, & [M_{+1}^{(n)}, M_{-1}^{(n)}] &= -\eta_{11} M_{+-}\,, & \nonumber\\
[M_{+-}, P_\pm^{(n)}] &= \pm\eta_{+-} P_\pm^{(n)}\,, & [M_{\pm 1}^{(n)}, P_\mp^{(n)}] &= -\eta_{+-} P_1\,, & [M_{\pm 1}^{(n)}, P_1] &= \eta_{11} P_\pm^{(n)}\,,
\end{align}
where $\eta$ is a Lorentzian metric with the non-zero components $\eta_{+-} = \eta_{-+} = -\eta_{11} = n$. For brevity, the commutation relations of generators with different $n$ (analogous to \eqref{eq:22.03}) are not shown. If we now rescale $M_{\pm 1}^{(n)} \mapsto c\, M_{\pm 1}^{(n)} = \widetilde{M}_{\pm 1}^{(n)}$, $P_1 \mapsto c\, P_1 = \widetilde{P}_1$ (for all $n \in \mathbb{N}$) and take the limit $c \to 0$, it allows to recover the CBMS${}_3$ algebra \eqref{BMS3C}. Similarly, rescaling $M_{\pm 1}^{(n)} \mapsto c^{-1} M_{\pm 1}^{(n)} = \widehat{M}_{\pm 1}^{(n)}$, $P_\pm^{(n)} \mapsto c^{-1} P_\pm^{(n)} = \widehat{P}_\pm^{(n)}$ (for all $n \in \mathbb{N}$) and taking the limit $c \to \infty$ allows to recover the GBMS${}_3$ algebra \eqref{BMS3G}. Let us stress that these contractions (in the sense of their meaning in terms of Poincar\'{e} embeddings) cannot be seen as a BMS-generalization of the Carrollian or Galilean contraction -- the rescaled Poincar\'{e} generators do not describe boosts and time translation, or boosts and spatial translations, respectively. We may call them ``quasi-Carrollian'' and ``quasi-Galilean''. %However, we will see in the next section that such contractions generalize to 3+1 dimensions, while the contractions associated with the family of embeddings \eqref{eq:22.01} possibly do not. 

If the quasi-Carrollian contraction is applied to the stand-alone Poincar\'{e} algebra, leading to the contraction limit:
\begin{alignat}{3}\label{eq:4.1a}
[M_{+-}, \widetilde{M}_{\pm 1}] &= \pm \widetilde{M}_{\pm 1}\,, &\qquad [\widetilde{M}_{+1}, \widetilde{M}_{-1}] &= 0\,, & \nonumber\\
[M_{+-}, P_\pm] &= \pm P_\pm\,, &\qquad [\widetilde{M}_{\pm 1}, P_\mp] &= -\widetilde{P}_1\,, &\qquad [\widetilde{M}_{\pm 1}, \widetilde{P}_1] &= 0\,,
\end{alignat}
it can be shown to have an unexpected connection to certain our results \cite{Kowalski:2014dy,Trzesniewski:2018ey} (see also \cite{Trzesniewski:2023gs}) obtained for (2+1)d gravity in the Chern-Simons formulation. Namely, using an isomorphism given by
\begin{align}
\tilde M := M_{+-}\,, \qquad \tilde K_a := \frac{1}{\sqrt{2}} \left(\widetilde{M}_{+1} \pm P_+\right), \qquad \tilde T_0 := -\widetilde{P}_1\,, \qquad \tilde T_a := \frac{1}{\sqrt{2}} \left(P_- \mp \widetilde{M}_{-1}\right),
\end{align}
we recover the algebra (in the form given in \cite{Trzesniewski:2018ey}) that we derived by contracting the Lorentz subgroup in the local Iwasawa decomposition of (2+1)d de Sitter group. Particles coupled to Chern-Simons theory with such a contracted gauge group turn out to satisfy the equations of motion of a Carroll particle, while symmetries of the action are a deformation of Carroll symmetries \cite{Kowalski:2014dy}, which further justifies the name ``quasi-Carrollian'' introduced above. We should also note that the algebra \eqref{eq:4.1a} is not a kinematical algebra and hence it is not included in the classification of \cite{Bacry:1986cy}.

Finally, turning to the remaining real forms of $\mathfrak{B}_3$, type $\mathfrak{so}(2,1)$ and $\mathfrak{su}(1,1)|\mathfrak{su}(2)$, let us try to impose the same principle of defining the Carrollian/Galilean contractions that lead to the Carroll/Galilei algebra when restricted to any Poincar\'{e} subalgebra. As we mentioned in Section~\ref{sec:3.0}, the subalgebras are now parametrized by $n \in 2\mathbbm{N}$. However, since taking the brackets of two generators with odd values of $n$ leads to the generators with even values of $n$ (cf. \eqref{eq:22.03}), in order to avoid divergences of the considered contractions, we would have to rescale the appropriate generators with odd values of $n$ instead of even, contradicting our assumption.

\subsection{$\Lambda$-BMS${}_3$}\label{sec:4.1}

A special feature of BMS algebra in 2+1 dimensions is that it can be generalized to the symmetry algebra of an asymptotically (Anti-)de Sitter spacetime, with the brackets ($n,m \in \mathbb{Z}$)
\begin{align}\label{3dalg}
[l_n, l_m] &= (n - m)\, l_{n + m}\,, \qquad [l_n, T_m] =  (n - m)\, T_{n + m}\,, \nonumber\\
[T_n, T_m] &= -\Lambda\, (n - m)\, l_{n + m}\,.
\end{align}
In general, it should be treated as a complex algebra known as $\Lambda$-BMS${}_3$. Depending on the sign of $\Lambda$, $\Lambda$-BMS${}_3$ can be equipped with one of several real structures, which allows to embed into it the (2+1)d de Sitter or Anti-de Sitter algebra (see below). The $\Lambda \rightarrow 0$ contraction limit of \eqref{3dalg} recovers the usual BMS${}_3$ algebra $\mathfrak{B}_3$, i.e. \eqref{BMS3}. 

Actually, transforming the brackets \eqref{3dalg} by an isomorphism (which is complex for $\Lambda > 0$)
\begin{align}\label{cwiso}
L_n := \frac{1}{2} \left(l_n + \frac{1}{\sqrt{-\Lambda}}\, T_n\right), \qquad \bar L_n := \frac{1}{2} \left(l_n - \frac{1}{\sqrt{-\Lambda}}\, T_n\right)
\end{align}
shows that the $\Lambda$-BMS${}_3$ algebra is equivalent to the direct sum of two copies of Witt algebra, $\mathfrak{W} \oplus \mathfrak{W}$:
\begin{align}\label{cwalg}
[L_n, L_m] = (n - m)\, L_{n + m}\,, \qquad [\bar L_n, \bar L_m] = (n - m)\, \bar L_{n + m}\,, \qquad [L_n, \bar L_m] = 0\,.
\end{align}
It is then easy to see that there exist infinitely many embeddings of the $\mathfrak{o}(4,\mathbb{C})$ algebra into $\Lambda$-BMS${}_3$, given by any subalgebra of the form
\begin{align}\label{embed1}
\mathfrak{o}(4, \mathbb{C}) \cong {\rm span}\{L_0,  L_{\pm n}, \bar L_0, \bar L_{\pm n}\} \subset \mathfrak{W} \oplus \mathfrak{W}\,, \quad n \in \mathbb{N}\,, %\\ L_n \rightarrow \frac{L_n}{n}\,, \qquad \bar L_n \rightarrow \frac{\bar L_n}{n}\,.
\end{align}
up to a rescaling of $L_n$, $\bar L_n$ by $1/n$. In terms of the generators $l_n$, $T_n$, these embeddings are given by
\begin{align}\label{embed2}
\mathfrak{o}(4, \mathbb{C}) \cong {\rm span}\{l_0, l_{\pm n}, T_0, T_{\pm n}\} \subset \mathfrak{W} \oplus \mathfrak{W}\,, \quad n \in \mathbb{N}\,, %\\ l_n \rightarrow \frac{l_n}{n}\,, \qquad \sqrt{-\Lambda} \rightarrow n \sqrt{-\Lambda}\,.
\end{align}
where (as one could expect due to the existence of $\Lambda \to 0$ contraction limit) the generators are the exact counterparts of those that span the embeddings of $\mathfrak{iso}(2,1)$ into $\mathfrak{B}_3$. 

As a consequence, possible real structures on the $\Lambda$-BMS${}_3$ algebra can be obtained by extending (non-compact) real forms of $\mathfrak{o}(4,\mathbb{C}) \cong \mathfrak{sl}(2,\mathbb{C}) \oplus \mathfrak{sl}(2,\mathbb{C})$ (cf. \cite{Borowiec:2017ag,Kowalski:2020qs}). Namely, we find the following real forms of $\Lambda$-BMS${}_3$:
\begin{itemize}
    \item[i)] type $\mathfrak{sl}(2,\mathbb{R}) \oplus \mathfrak{sl}(2,\mathbb{R})$, for which $\ L_n^* = -L_n\,, \quad \bar L_n^* = -\bar L_n\,, \quad n \in \mathbb{Z}\,$,
    \item[ii)] type $\mathfrak{su}(1,1) \oplus \mathfrak{su}(1,1)$, for which $\ L_n^* = L_{-n}\,, \quad \bar L_n^* = \bar L_{-n}\,, \quad n \in \mathbb{Z}\,$,
    \item[iii)] type $\mathfrak{sl}(2,\mathbb{R}) \oplus \mathfrak{su}(1,1)$, for which $\ L_n^* = -L_n\,, \quad \bar L_n^* = \bar L_{-n}\,, \quad n \in \mathbb{Z}\,$,
    \item[iv)] type $\mathfrak{so}(3,1)_{\rm a}$, for which $\ L_n^* = -\bar L_n\,, \quad n \in \mathbb{Z}\,$,
    \item[v)] type $\mathfrak{so}(3,1)_{\rm b}$, for which $\ L_n^* = \bar L_{-n}\,, \quad n \in \mathbb{Z}\,$.
\end{itemize}
%In particular, the generators $l_n$, $T_n$ satisfy the anti-Hermitian reality conditions in the case i) with $\Lambda < 0$ and in the case iv) with $\Lambda > 0$. 
Let us recall that $\mathfrak{so}(3,1)$ is the (2+1)d de Sitter algebra, while the real forms of $\mathfrak{o}(4,\mathbb{C})$ from the cases i-iii) are isomorphic to the (2+1)d Anti-de Sitter algebra $\mathfrak{so}(2,2)$; types $\mathfrak{so}(3,1)_{\rm a}$ and $\mathfrak{so}(3,1)_{\rm b}$ are two BMS-extensions of $\mathfrak{so}(3,1)$. The remaining real forms of $\mathfrak{o}(4,\mathbb{C})$ ($\mathfrak{o}(4)$, $\mathfrak{o}^\star(4)$ and $\mathfrak{o}'^\star(4)$) cannot be extended to $\Lambda$-BMS${}_3$ because it leads to the $\mathfrak{su}(2)$-type reality conditions on its subalgebra spanned by the generators $L_n$, which is incompatible with the Lie bracket on $\mathfrak{W}$ (as we already noted in our discussion of real forms of $\mathfrak{B}_3$). 

Based on this whole discussion, it is justified to apply to \eqref{3dalg} the same procedure of the Carrollian/Galilean contraction as we defined for the $\mathfrak{B}_3$ algebra in the basis $\{l_n,T_n;\, n \in \mathbb{Z}\}$. In the Carrollian case, the contraction leads to the brackets identical to \eqref{BMS3C} plus the additional non-trivial ones in the supertranslation sector:
\begin{align}\label{LBMS3C}
\big[T_{n}, \widetilde{T}_{0}\big] = -\Lambda\, n\, \tilde{l}_{n}\,, \qquad \big[T_{n}, T_{-n}\big] = -2\Lambda\, n\, l_{0}\,, \qquad \big[T_{n}, T_{m}\big] \to \infty\,,\ \ m \neq -n
\end{align}
but the divergence means that, surprisingly, the contraction limit does not exist. On the other hand, in the Galilean case, we obtain the brackets identical to \eqref{BMS3G} plus the following ones in the supertranslation sector:
\begin{align}\label{LBMS3G}
\big[\widehat{T}_{n}, T_{0}\big] = -\Lambda\, n\, \widehat{l}_{n}\,, \qquad \big[\widehat{T}_{n}, \widehat{T}_{-n}\big] = 0\,, \qquad \big[\widehat{T}_{n}, \widehat{T}_{m}\big] = 0\,,\ \ m \neq -n\,.
\end{align}
The algebra with such brackets may be called Galilei-$\Lambda$-BMS${}_3$. However, similarly to what we showed for the $\mathfrak{B}_3$ algebra, the interpretation of the above contractions depends on a chosen real form of $\Lambda$-BMS${}_3$. Let us now discuss that. 

It is quite obvious that the reality conditions for the generators $l_n$, $T_n$ in the $\mathfrak{su}(1,1)$-type real form of the $\mathfrak{B}_3$ algebra are the same as in the $\Lambda$-BMS${}_3$ real forms: type $\mathfrak{su}(1,1) \oplus \mathfrak{su}(1,1)$ if $\Lambda < 0$, or type $\mathfrak{so}(3,1)_{\rm b}$ if $\Lambda > 0$. Consequently, the formulae \eqref{eq:22.01} can also be used to describe a family of embeddings of either $\mathfrak{so}(2,2)$ or $\mathfrak{so}(3,1)$ into $\Lambda$-BMS${}_3$. Generalizing our approach applied in the case of $\mathfrak{B}_3$ in \eqref{BMS3C}, \eqref{BMS3G}, if we consider one of those two real forms of $\Lambda$-BMS${}_3$, the contractions leading to \eqref{LBMS3C}, \eqref{LBMS3G} can actually be inferred as the natural extension of the Carrollian/Galilean contractions of (Anti-)de Sitter algebra leading to \eqref{eq:11d}, \eqref{eq:11e}. 

Meanwhile, the $\mathfrak{sl}(2,\mathbb{R})$-type real form of $\mathfrak{B}_3$ is in agreement with the $\Lambda$-BMS${}_3$ real forms: type $\mathfrak{sl}(2,\mathbb{R}) \oplus \mathfrak{sl}(2,\mathbb{R})$ if $\Lambda < 0$, or type $\mathfrak{so}(3,1)_{\rm a}$ if $\Lambda > 0$. Defining a family of embeddings of either $\mathfrak{so}(2,2)$ or $\mathfrak{so}(3,1)$ into $\Lambda$-BMS${}_3$, given by the formulae identical to \eqref{eq:22.02}, we find that the contractions leading to \eqref{LBMS3C}, \eqref{LBMS3G} have to be interpreted as quasi- Carrollian/Galilean, in terms of the $\Lambda$-BMS${}_3$-generalization of the basis \eqref{eq:4.1}. The latter includes additional non-vanishing brackets
\begin{align}
[P_\pm^{(n)},P_1] = -\Lambda\, \eta_{11} M_{\pm 1}^{(n)}\,, \qquad [P_+^{(n)}, P_-^{(n)}] = -\Lambda\, \eta_{11} M_{+-}\,,
\end{align}
as well as the ones for the $P_\pm^{(n)}$ generators with different $n$. Naturally, the calculations made in this basis confirm that the (quasi-)Carrollian contraction limit \eqref{LBMS3C} is divergent. 

Finally, the $\mathfrak{sl}(2,\mathbb{R}) \oplus \mathfrak{su}(1,1)$-type reality conditions are given by more complicated expressions:
\begin{align}
l_n^* = \frac{1}{2} \left(l_{-n} - l_n\right) - \frac{1}{2 \sqrt{-\Lambda}} \left(T_{-n} + T_n\right), \qquad T_n^* = \frac{1}{2} \left(T_{-n} - T_n\right) - \frac{\sqrt{-\Lambda}}{2} \left(l_{-n} + l_n\right),
\end{align}
which are different than for any real form of $\mathfrak{B}_3$. They are also inconsistent with the rescalings applied in the Carrollian contraction, due to the mixing of $l_n$'s with $T_n$'s. Therefore, this case seems to be of lesser interest and we will not delve deeper into it. 

One might still imagine that the Carrollian counterpart of $\Lambda$-BMS${}_3$ exists but cannot be reached by the contraction procedures considered here. In order to find such a hypothetical Carroll-$\Lambda$-BMS${}_3$ algebra, we could assume that the well-known isomorphism between Poincar\'{e} and AdS-Carroll algebras generalizes to the BMS case. Namely, based on the form of Poincar\'{e} embeddings \eqref{eq:22.01} into the $\mathfrak{B}_3$ algebra, let us define the map that transforms the standard basis of the latter $\{l_n,T_n;\, n \in \mathbb{Z}\}$ to a new basis $\{l'_n,T'_n;\, n \in \mathbb{Z}\}$:
\begin{align}\label{eq:12x}
l_k \mapsto {\rm sgn}(k)\, i \sqrt{-\Lambda}^{-1} T'_k\,, \qquad T_k \mapsto {\rm sgn}(k)\, i \sqrt{-\Lambda}\, l'_k\,, \qquad l_0 \mapsto l'_0\,, \qquad T_0 \mapsto T'_0
\end{align}
($k \neq 0$), so that it reduces to
\begin{align}\label{eq:12}
K_a \mapsto \sqrt{-\Lambda}^{-1} {\cal T}_a\,, \quad P_a \mapsto -\sqrt{-\Lambda}\, Q_a\,, \quad J \mapsto R\,, \quad P_0 \mapsto {\cal T}_0
\end{align}
for each Poincar\'{e} subalgebra (cf. \eqref{eq:10}, \eqref{eq:11d}). It turns out that the brackets of $\mathfrak{B}_3$ in the new basis are identical to \eqref{BMS3C}, \eqref{LBMS3C} with two exceptions: $[T'_n, T'_m]$ is no longer divergent but $[l'_n, T'_m]$ differs from its counterpart given in \eqref{BMS3C}. It follows that the map \eqref{eq:12x} does not provide us with what one could call the Carroll-$\Lambda$-BMS${}_3$ algebra.

However, this does not necessarily mean that a Carrollian spacetime with $\Lambda < 0$ has no consistent algebra of asymptotic symmetries. For comparison, a recent analysis of the Carroll limit of asymptotically Anti-de Sitter spacetime in 3+1 dimensions has shown \cite{Perez:2022at} that the corresponding algebra is actually $\mathfrak{B}_4$ (which reduces to the standard BMS${}_4$, without superrotations, if one sets more restrictive boundary conditions); precisely speaking, such is the case for only one of the two possible Carroll limits of General Relativity, the so-called magnetic one, while the electric limit leads to an inconsistency.

\section{Contractions of (extended) BMS${}_4$ algebra $\mathfrak{B}_4$}\label{sec:5.0}

The extended BMS algebra in 3+1 dimensions (introduced by Barnich and Troessaert \cite{Barnich:2010sd}), which we will denote $\mathfrak{B}_4$, has the brackets
\begin{align}\label{bms}
[L_n, L_m] &= (n - m)\, L_{n + m}\,, & [\bar L_n, \bar L_m] &= (n - m)\, \bar L_{n + m}\,, & [L_n, \bar L_m] &= 0\,, \nonumber\\
[L_k, T_{nm}] &= \left(\frac{k + 1}{2} - n\right) T_{n + k,m}\,, & [\bar L_k, T_{nm}] &= \left(\frac{k + 1}{2} - m\right) T_{n,m + k}\,, & [T_{nm}, T_{n'm'}] &= 0
\end{align}
in terms of the generators of superrotations $L_n$, $\bar L_n$ and supertranslations $T_{nm}$, where $n, m \in \mathbb{Z}$. Similarly to $\mathfrak{B}_3$ and $\Lambda$-BMS${}_3$, $\mathfrak{B}_4$ is in general a complex algebra. Its subalgebra spanned by $\{L_n,\bar L_n;\, n \in \mathbb{Z}\}$ is obviously isomorphic to $\Lambda$-BMS${}_3$, cf. \eqref{cwalg}. There are four real structures that one can introduce here as generalizations of the $\Lambda$-BMS${}_3$ real forms:
\begin{itemize}
	\item[i)] type $\mathfrak{sl}(2,\mathbb{R}) \oplus \mathfrak{sl}(2,\mathbb{R})$, for which $\ L_{n}^* = -L_{n}\,, \quad \bar L_{n}^* = -\bar L_{n}\,, \quad T_{nm}^* = -T_{nm}\,$,
	\item[ii)] type $\mathfrak{su}(1,1) \oplus \mathfrak{su}(1,1)$, for which $\ L_{n}^* = L_{-n}\,, \quad \bar L_{n}^* = \bar L_{-n}\,, \quad T_{nm}^* = -T_{1-n,1-m}\,$,
	\item[iii)] type $\mathfrak{so}(3,1)_{\rm a}$, for which $\ L_n^* = -\bar L_{n}\,, \quad T_{nm}^* = T_{mn}\,$,
	%\begin{align}\label{eq:50.01}
	%L_n^* = -\bar L_{n}\,, \qquad T_{nm}^* = T_{mn}\,,
	%\end{align}
	\item[iv)] type $\mathfrak{so}(3,1)_{\rm b}$, for which $\ L_n^* = \bar L_{-n}\,, \quad T_{nm}^* = T_{1-m,1-n}\,$.
	%\begin{align}\label{eq:50.01a}
	%L_n^* = \bar L_{-n}\,, \qquad T_{nm}^* = T_{1-m,1-n}\,.
	%\end{align}
\end{itemize}
The real form of the $\mathfrak{sl}(2,\mathbb{R}) \oplus \mathfrak{su}(1,1)$-type, for which $L_{n}^* = -L_{n}\,,\ \bar L_{n}^* = \bar L_{-n}$, does not extend to the supertranslation sector. 

The real forms of types $\mathfrak{so}(3,1)_{\rm a}$ and $\mathfrak{so}(3,1)_{\rm b}$ are physically the most interesting ones, since then the subalgebra spanned by $\{L_n,\bar L_n;\, n \in \mathbb{Z}\}$ can be seen as the BMS generalization of Lorentz algebra $\mathfrak{so}(3,1)$. We will restrict here to the type $\mathfrak{so}(3,1)_{\rm a}$ and express the $\mathfrak{B}_4$ algebra in a basis consisting of the following anti-Hermitian generators
%We will first consider the type $\mathfrak{so}(3,1)_{\rm b}$ and express the $\mathfrak{B}_4$ algebra in a basis consisting of anti-Hermitian generators
%We now may turn to the real form type $\mathfrak{so}(3,1)_{\rm a}$ and express the $\mathfrak{B}_4$ algebra in a basis consisting of anti-Hermitian generators
\begin{alignat}{3}
R_n &:= L_n + \bar L_n\,, &\qquad \bar R_n &:= -i\, (L_n - \bar L_n)\,, \nonumber\\
S_{nm} &:= \frac{i}{2}\, (T_{nm} + T_{mn})\,, &\qquad A_{nm} &:= \frac{1}{2}\, (T_{nm} - T_{mn})
\end{alignat}
(let us note that $S_{nm}$'s are symmetric under the exchange of indices $n \leftrightarrow m$, while $A_{nm}$'s are antisymmetric and hence there are no such generators with $n = m$; however, for brevity, the symbol $A_{nn}$ is implicitly used below in the sense of $A_{nn} = 0$), so that the brackets \eqref{bms} become
\begin{align}\label{4dembed1}
[R_n, R_m] &= (n - m)\, R_{n+m}\,, \qquad [\bar R_n, \bar R_m] = -(n - m)\, R_{n+m}\,, \nonumber\\
[R_n, \bar R_m] &= (n - m)\, \bar R_{n+m}\,,
\nonumber\\
[R_k, S_{nm}] &= \left(\frac{k + 1}{2} - n\right) S_{n + k, m} + \left(\frac{k + 1}{2} - m\right) S_{n, m + k}\,, \nonumber\\
[\bar R_k, S_{nm}] &= \left(\frac{k + 1}{2} - n\right) A_{n + k, m} - \left(\frac{k + 1}{2} - m\right) A_{n, m + k}\,, \nonumber\\
[R_k, A_{nm}] &= \left(\frac{k + 1}{2} - n\right) A_{n + k, m} + \left(\frac{k + 1}{2} - m\right) A_{n, m + k}\,, \nonumber\\
[\bar R_k, A_{nm}] &= -\left(\frac{k + 1}{2} - n\right) S_{n + k, m} + \left(\frac{k + 1}{2} - m\right) S_{n, m + k}\,.
\end{align}
The new basis makes it evident that the algebraic structure of $\mathfrak{B}_4$ (albeit infinite-dimensional) is analogous to that of a kinematical (Lie) algebra. It follows from the definition \cite{Bergshoeff:2023ar} that a kinematical algebra contains the subalgebra of rotations, that the generators of boosts and spatial translations transform under rotations as vectors, and that the generator of time translations transforms under rotations as a scalar. Accordingly, \eqref{4dembed1} contains the rotation-like subalgebra spanned by the generators $R_n$, which act on $\bar R_n$, $S_{pq}$ and $A_{pq}$ in a vector-like way; the only missing piece seems to be an analogue of the time translation generator. If we narrow down our analogy to Poincar\'{e} algebra, we can identify $\bar R_n$'s as corresponding to the boost generators and $S_{pq}$'s -- to the spatial translation generators, while $A_{pq}$'s fill the spot of the time translation generator, with a caveat that they do not commute with $R_n$'s (in particular, let us note that if $p,q \in \{-n,n\}$, there are three generators $S_{pq}$ and only one $A_{pq}$; however, this does not allow to construct an embedding of $\mathfrak{iso}(3,1)$ into $\mathfrak{B}_4$). Consequently, it turns out that one can perform two contractions of $\mathfrak{B}_4$ similar to the Carrollian and Galilean contraction of Poincar\'{e} algebra. As we will see, however, these contractions are not generalizations of the corresponding ones for Poincar\'{e} algebra and hence we call them quasi- Carrollian/Galilean. 

The quasi-Carrollian contraction consists in the rescalings $\bar R_n \mapsto c\, \bar R_n$ and $A_{nm} \mapsto c\, A_{nm}$, $\forall n,m \in \mathbb{Z}$, and then taking the limit $c \to 0$, which leads to the algebra:
\begin{align}\label{4dembed1C}
[R_n, R_m] &= (n - m)\, R_{n+m}\,, \qquad [R_n, \bar R_m] = (n - m)\, \bar R_{n+m}\,, \qquad [\bar R_n, \bar R_m] = 0\,, \nonumber\\
[R_k, S_{nm}] &= \left(\frac{k + 1}{2} - n\right) S_{n + k, m} + \left(\frac{k + 1}{2} - m\right) S_{n, m + k}\,, \nonumber\\
[\bar R_k, S_{nm}] &= \left(\frac{k + 1}{2} - n\right) A_{n + k, m} - \left(\frac{k + 1}{2} - m\right) A_{n, m + k}\,, \nonumber\\
[R_k, A_{nm}] &= \left(\frac{k + 1}{2} - n\right) A_{n + k, m} + \left(\frac{k + 1}{2} - m\right) A_{n, m + k}\,, \qquad [\bar R_k, A_{nm}] = 0\,.
\end{align}
Meanwhile, the quasi-Galilean contraction of \eqref{4dembed1} is performed by rescaling $\bar R_n \mapsto c^{-1} \bar R_n$ and $S_{nm} \mapsto c^{-1} S_{nm}$, $\forall n,m \in \mathbb{Z}$, and then taking the limit $c \to \infty$, which gives us the algebra:
\begin{align}\label{4dembed1G}
[R_n, R_m] &= (n - m)\, R_{n+m}\,, \qquad [R_n, \bar R_m] = (n - m)\, \bar R_{n+m}\,, \qquad [\bar R_n, \bar R_m] = 0 \,, \nonumber\\
[R_k, S_{nm}] &= \left(\frac{k + 1}{2} - n\right) S_{n + k, m} + \left(\frac{k + 1}{2} - m\right) S_{n, m + k}\,, \qquad [\bar R_k, S_{nm}] = 0\,, \nonumber\\
[R_k, A_{nm}] &= \left(\frac{k + 1}{2} - n\right) A_{n + k, m} + \left(\frac{k + 1}{2} - m\right) A_{n, m + k}\,, \nonumber\\
[\bar R_k, A_{nm}] &= -\left(\frac{k + 1}{2} - n\right) S_{n + k, m} + \left(\frac{k + 1}{2} - m\right) S_{n, m + k}\,.
\end{align}
The use of terms ``quasi-Carrollian'' and ``quasi-Galilean'' can be justified by referring to the analogy between the $\mathfrak{B}_4$ and Poincar\'{e} algebra discussed earlier. We observe that both algebras \eqref{4dembed1C} and \eqref{4dembed1G} have commuting boost-like generators $\bar R_n$, as well as $\bar R_n$ commuting with $A_{nm}$'s in the first case and $\bar R_n$ commuting with $S_{nm}$'s in the second case, while the remaining brackets are not changed with respect to \eqref{4dembed1}. If $A_{nm}$, $S_{nm}$ are interpreted as time and spatial ``translation-like'' generators (as we already postulated for $\mathfrak{B}_4$), respectively, the above-mentioned commutation relations have exactly the same structure as for Carroll and Galilei algebras, respectively. Therefore, we will call the algebra defined by \eqref{4dembed1C} quasi-Carroll-BMS${}_4$ and the algebra defined by \eqref{4dembed1G} -- quasi-Galilei-BMS${}_4$. 

Let us now recall that one can identify the Poincar\'{e} algebra $\mathfrak{iso}(3,1)$ with a maximal finite-dimensional subalgebra of $\mathfrak{B}_4$. Similarly as in the case of 2+1 dimensions, there exist infinitely many such subalgebras, spanned by 10 generators that satisfy the commutation relations generalizing \eqref{eq:22.00} to 4 dimensions:
\begin{alignat}{6}\label{eq:P4}
[J_a,J_b] &= \eta_{00} \epsilon_{abc} J_c\,, &\qquad [K_a,K_b] &= - \eta_{00} \epsilon_{abc} J_c\,, &\qquad 
[J_a,K_b] &= \eta_{00} \epsilon_{abc} K_c\,, &\qquad [J_a,P_0] &= 0\,, \nonumber\\ 
[J_a,P_b] &= \eta_{00} \epsilon_{abc} P_c\,, &\qquad [K_a,P_0] &= \eta_{00} P_a\,, &\qquad [K_a,P_b] &= -\eta_{ab} P_0\,, &\qquad [P_\mu,P_\nu] &= 0\,.
\end{alignat}
(This form of the brackets can be recovered from \eqref{Poincare} by taking $M_{ab} = \epsilon_{abc} J_c$, $M_{a0} = K_a$.) The above-mentioned subalgebras correspond to a family of embeddings of $\mathfrak{iso}(3,1)$ into $\mathfrak{B}_4$, parametrized by $n \in 2\mathbb{N} - 1$ (cf. \cite{Borowiec:2021bs}):
\begin{alignat}{4}\label{eq:5.1}
J_1^{(n)} &= \frac{1}{2} (\bar R_{-n} - \bar R_n)\,, &\qquad J_2^{(n)} &= \frac{1}{2} (R_{-n} + R_n)\,, &\qquad J_3 &= -\bar R_0\,, \nonumber\\
K_1^{(n)} &= \frac{1}{2} (R_n - R_{-n})\,, &\qquad K_2^{(n)} &= \frac{1}{2} (\bar R_n + \bar R_{-n})\,, &\qquad K_3 &= R_0\,, \nonumber\\
P_0^{(n)} &= \frac{1}{2} (S_{qq} + S_{pp})\,, &\qquad P_3^{(n)} &=  \frac{1}{2} (S_{qq} - S_{pp})\,, &\qquad P_1^{(n)} &= S_{pq}\,, \qquad P_2^{(n)} = A_{pq}\,, %= -A_{qp}
\end{alignat}
where the indices $p = (1 + n)/2$, $q = (1 - n)/2$, so that the new generators satisfy the brackets \eqref{eq:P4} with the metric $\eta = n\, {\rm diag}(1,-1,-1,-1)$. The Poincar\'{e} generators are anti-Hermitian ($X^* = -X$) if we impose the reality conditions of the type $\mathfrak{so}(3,1)_{\rm a}$ on the $\mathfrak{B}_4$ algebra. 

Such an embedding of the Lorentz algebra is well-defined also for even $n \neq 0$, while the same is not true for the translation generators. Consequently, in contrast to what happens in 2+1 dimensions, the union of all embeddings from the family does not correspond to a basis of the full $\mathfrak{B}_4$ algebra. Furthermore, the brackets between $J_a^{(n)}$ and $K_b^{(m)}$ with odd parameters $n,m$ are given by generators with even parameters $n+m$ and $n-m$ and hence the union of these embeddings is not even a subalgebra. Taking all this into account, we conclude that the embeddings in the 3+1-dimensional case do not allow us to straightforwardly generalize the contractions of Poincar\'{e} to BMS algebra; conversely, performing the contractions of BMS is independent from embedding Poincar\'{e} algebra into it. In particular, the form of translation generators in \eqref{eq:5.1} does not preserve our interpretation of $A_{nm}$'s and $S_{nm}$'s as the generalizations of time and spatial translation generators, respectively, while the rotation and boost generators mix $R_n$'s and $\bar R_n$'s. However, this apparent inconsistency will find a resolution below. 

Let us consider a different picture. Choosing two light-like vectors in flat spacetime, an outgoing $\tau_+ = 1/\sqrt{2}\, (1,0,0,1)$ and incoming $\tau_- = 1/\sqrt{2}\, (1,0,0,-1)$, one can perform the 2+2 spacetime decomposition and introduce light-cone (a.k.a. light-front) generators of Poincar\'{e} algebra \eqref{Poincare},
\begin{align}
M_{\pm a} &= \tau_{\pm}^{\mu} M_{\mu a} = \frac{1}{\sqrt{2}} \left(M_{0a}\pm M_{3a}\right), \qquad M_{+-} = \tau_+^\mu\tau_-^\nu M_{\mu\nu} = M_{30}\,, \nonumber\\
P_\pm &= \frac{1}{\sqrt{2}} \left(P_0 \pm P_3\right),
\end{align}
so that, for a general metric $\eta$, their commutation relations have the form
\begin{alignat}{3}\label{eq:50.02}
& [M_{+a},M_{-b}] = -\eta_{+-} M_{ab} - \eta_{ab} M_{+-}\,, \qquad && [M_{\pm a},M_{\pm b}] = 0\,, & 
\nonumber\\
& [M_{\pm a},M_{bc}] = \eta_{ab} M_{\pm c} - \eta_{ac} M_{\pm b}\,, \qquad && [M_{+-},M_{\pm a}] = \pm\eta_{+-} M_{\pm a}\,, 
\nonumber\\
& [M_{\pm a},P_{\mp}] = -\eta_{+-} P_{a}\,, \qquad && [M_{\pm a},P_{\pm}] = [M_{+-},P_{a}] = 0\,, & 
\nonumber\\
& [M_{\pm a},P_{b}] = \eta_{ab} P_{\pm}\,, \qquad && [M_{+-},P_{\pm}] = \pm \eta_{+-} P_{\pm} & 
%\nonumber\\
%&[M_{ab},P_{c}] = \eta_{bc} P_{a} - \eta_{ac} P_{b}\,, \qquad && 
\end{alignat}
($a,b,c = 1,2$). Similarly to the 2+1-dimensional case \eqref{eq:4.1}, $M_{\pm 1}$, $M_{\pm 2}$ are actually generators of null rotations (transformations generated by $M_{\pm a}$ leave invariant the null directions $\tau_\pm$, as well as the spatial directions $x_2, x_1$ if $a = 1,2$, respectively) and $P_{\pm}$ are generators of translations along the null directions $\tau_\pm$, respectively. 

Transforming the image of a given embedding \eqref{eq:5.1} to such a light-cone basis, we obtain:
\begin{align}
M_{\pm 1}^{(n)} &= \frac{1}{\sqrt{2}} \left(-K_1^{(n)} \pm J_2^{(n)}\right) = \pm\frac{1}{\sqrt{2}}\, R_{\mp n}\,, \qquad M_{+-} = K_3 = R_0\,, \nonumber\\
M_{\pm 2}^{(n)} &= \frac{1}{\sqrt{2}} \left(-K_2^{(n)}\mp J_1^{(n)}\right) = -\frac{1}{\sqrt{2}}\, \bar R_{\mp n}\,, \qquad M_{12} = J_3 = -\bar R_0\,, \nonumber\\
P_+^{(n)} &= \frac{1}{\sqrt{2}}\, S_{qq}\,, \qquad P_-^{(n)} = \frac{1}{\sqrt{2}}\, S_{pp}\,, \qquad P_1^{(n)} = S_{pq}\,, \qquad P_2^{(n)} = A_{pq}
\end{align}
and then the brackets \eqref{eq:50.02} are satisfied with the metric components $\eta_{+-} = \frac{1}{2} (\eta_{00} - \eta_{33}) = -\eta_{aa} = n$ and $\eta_{++} = \eta_{--} = \eta_{ab} = 0$, $a \neq b$. 

From this point of view, the quasi-Carrollian contraction involves rescaling of the generators generalizing translations along the spatial direction $x_2$, while for the quasi-Galilean contraction one needs to rescale the generators generalizing translations along the null directions $\tau_\pm$ and the spatial direction $x_1$; and, in both cases, rescale the generators generalizing null rotations that leave invariant the directions $\tau_\pm$ and $x_1$, as well as the generator of rotations in the $x_1 x_2$ plane. This can now be compared with contractions of the $\mathfrak{B}_3$ algebra performed in the basis \eqref{eq:4.1}. Indeed, in that case, the quasi-Carrollian contraction also involved rescaling of a spatial translation generator, while the quasi-Galilean contraction -- rescaling of the null translation generators, and for both contractions we also rescaled the null rotation generators (while there is no 2+1-dimensional counterpart of the rotation generator $M_{12}$). The difference is that the limits of such contractions in 2+1 dimensions are actually given by the Carroll-BMS${}_3$ and Galilei-BMS${}_3$ algebras (which are also derivable via the ``non-quasi'' Carrollian and Galilean contractions, performed in \eqref{BMS3C} and \eqref{BMS3G}), while we do not obtain algebras that could be described as Carroll-BMS${}_4$ and Galilei-BMS${}_4$. 

%Using the terminology from field theory \cite{Bengtsson:1983cn}, we can also say that the quasi-Carroll-BMS${}_4$ algebra \eqref{4dembed1C} contains the kinematical subalgebra (not to be confused with the concept of kinematical Lie algebras, to which we refer elsewhere) spanned by the generators $\{R_n, S_{pq};\, n \in 2\mathbb{N} - 1\}$ and the dynamical subalgebra spanned by $\{\bar R_n, A_{pq};\, n \in 2\mathbb{N} - 1\}$, while the kinematical subalgebra of the quasi-Galilei-BMS${}_4$ algebra \eqref{4dembed1G} is spanned by the generators $\{R_n, A_{pq};\, n \in 2\mathbb{N} - 1\}$ and its dynamical subalgebra by $\{\bar R_n, S_{pq};\, n \in 2\mathbb{N} - 1\}$.
%based on the family of $\mathfrak{iso}(2,1)$ embeddings \eqref{eq:22.02}
%associated with the other family of $\mathfrak{iso}(2,1)$ embeddings \eqref{eq:22.01}

\section{Conclusions}\label{sec:6.0}

The aim of this paper was to try to define the notions of Carrollian and Galilean contractions of BMS algebra in 3+1 and 2+1 dimensions, as well as $\Lambda$-BMS in 2+1 dimensions. 

We showed that both types of contractions can be extended to the BMS${}_3$ algebra, leading to the BMS counterparts of Carroll and Galilei algebras, called Carroll-BMS${}_3$ and Galilei-BMS${}_3$. This is achieved by using a family of embeddings of Poincar\'{e} algebra $\mathfrak{iso}(2,1)$ into $\mathfrak{B}_3$, which allows us to decompose the latter into (overlapping) subalgebras isomorphic to $\mathfrak{iso}(2,1)$. We adopt the condition that the contraction limit of each such subalgebra should be isomorphic to Carroll/Galilei algebra and find that it can be satisfied if the considered embeddings are associated with a particular real form (type $\mathfrak{su}(1,1)$) of $\mathfrak{B}_3$. On the other hand, CBMS${}_3$ and GBMS${}_3$ can also be recovered by starting with a family of embeddings associated with the other real form (type $\mathfrak{sl}(2,\mathbb{R})$). The contractions of BMS${}_3$ that we perform in such a case are not equivalent to the Carrollian/Galilean contractions when restricted to the corresponding Poincar\'{e} subalgebras, hence we call them quasi- Carrollian/Galilean. Moreover, as we discussed, they turn out to have an interesting connection to a particular contraction considered for the gauge group of the Chern-Simons theory describing (2+1)d gravity with $\Lambda > 0$. 

Each of the above contraction procedures can be straightforwardly generalized to the $\Lambda$-BMS${}_3$ algebra: Carrollian/Galilean (for the real form type $\mathfrak{su}(1,1) \oplus \mathfrak{su}(1,1)$ or $\mathfrak{so}(3,1)_{\rm b}$) or quasi- Carrollian/Galilean (for the real form type $\mathfrak{sl}(2,\mathbb{R}) \oplus \mathfrak{sl}(2,\mathbb{R})$ or $\mathfrak{so}(3,1)_{\rm a}$). We find that only the Galilean contraction limit is well-defined in these cases, which seems a bit puzzling, although the previously mentioned results of \cite{Perez:2022at} shed some light here. 

BMS algebra in 3+1 dimensions ($\mathfrak{B}_4$) is more problematic because the embeddings of Poincar\'{e} algebra $\mathfrak{iso}(3,1)$ do not cover the whole algebra and hence do not provide a framework for extending the contractions of $\mathfrak{iso}(3,1)$ to $\mathfrak{B}_4$. As an alternative method, we observe the analogy between the structures of these two algebras and find that (at least, considering the $\mathfrak{so}(3,1)_{\rm a}$-type real form) it is only possible to perform the so-called quasi-Carrollian and quasi-Galilean contractions of $\mathfrak{B}_4$, which lead to the quasi-Carroll-BMS${}_4$ and quasi-Galilei-BMS${}_4$ algebras. The structure of these algebras is superficially reminiscent of Carroll and Galilei algebras but does not agree with the embeddings of $\mathfrak{iso}(3,1)$ into $\mathfrak{B}_4$. However, when expressed in an appropriate basis, the contractions turn out to be a straightforward generalization of the quasi-Carrollian and quasi-Galilean contractions of $\mathfrak{B}_3$. %which further confirms that the former are not defined in an arbitrary manner. 

The contractions of BMS algebras are of interest on their own, but their construction is also the first step in investigations of their quantum deformations. This can be done similarly to the construction presented in \cite{Borowiec:2019ky,Borowiec:2021bs,Borowiec:2021ds} and such a generalization of the results obtained for quantum deformations of finite-dimensional symmetry algebras \cite{Trzesniewski:2023qi,Ballesteros:2020ts} will be the subject of our follow-up paper.

\section*{Acknowledgments}

TT and AB are supported by the National Science Center, project no. UMO-2022/45/B/ST2/01067. For JKG, this work was supported by funds provided by the National Science Center, project no. 2019/33/B/ST2/00050. The authors are grateful for discussions with Glenn Barnich, especially during his visit at the University of Wroc\l aw.


\begin{thebibliography}{99}

\bibitem{Arnowitt:1962hi}
R.~L.~Arnowitt, S.~Deser and C.~W.~Misner, 
{\it The Dynamics of general relativity,} in {\it Gravitation: an introduction to current research}, 
L. Witten, ed. (Wiley, New York, 1962) republished in
Gen.\ Rel.\ Grav.\ {\bf 40} (2008) 1997 
[arXiv:gr-qc/0405109].

\bibitem{Regge:1974zd}
T.~Regge and C.~Teitelboim, 
{\it Role of Surface Integrals in the Hamiltonian Formulation of General Relativity}, 
Annals Phys.\ {\bf 88} (1974) 286.

\bibitem{Bondi:1962gm}
H.~Bondi, M.~G.~J.~van der Burg and A.~W.~K.~Metzner, 
{\it Gravitational waves in general relativity, VII. Waves from axi-symmetric isolated systems}, 
Proc.\ Roy.\ Soc.\ Lond. A {\bf 269} (1962) 21.

\bibitem{Sachs:1962gs}
R.~K.~Sachs, 
{\it Gravitational waves in general relativity, VIII. Waves in asymptotically flat space-times}, 
Proc.\ Roy.\ Soc.\ Lond. A {\bf 270} (1962) 103.

\bibitem{Strominger:2017zoo}
A.~Strominger, 
{\it Lectures on the Infrared Structure of Gravity and Gauge Theory}, 
arXiv:1703.05448 [hep-th].

\bibitem{Barnich:2010sd}
G.~Barnich and C.~Troessaert, 
{\it Symmetries of Asymptotically Flat Four-Dimensional Spacetimes at Null Infinity Revisited}, 
Phys.\ Rev.\ Lett.\ {\bf 105} (2010) 111103 
[arXiv:0909.2617 [gr-qc]].

\bibitem{Barnich:2010eb}
G.~Barnich and C.~Troessaert, 
{\it Aspects of the BMS/CFT correspondence}, 
JHEP {\bf 05} (2010) 062 
[arXiv:1001.1541 [hep-th]].

\bibitem{Geiller:2024cs}
M.~Geiller, 
{\it Celestial $w_{1+\infty}$ charges and the subleading structure of asymptotically-flat spacetimes}, 
arXiv:2403.05195 [hep-th].

\bibitem{Henneaux:2018cst}
M.~Henneaux and C.~Troessaert, 
{\it BMS Group at Spatial Infinity: the Hamiltonian (ADM) approach}, 
JHEP {\bf 03} (2018) 147 
[arXiv:1801.03718 [gr-qc]].

\bibitem{Brocki:2021ieh}
L.~Brocki and J.~Kowalski-Glikman, 
{\it On symmetries and charges at spatial infinity}, 
arXiv:2109.06642 [gr-qc].

\bibitem{Grumiller:2020sr}
D.~Grumiller, A.~P\'{e}rez, M.~M.~Sheikh-Jabbari, R.~Troncoso and C.~Zwikel, 
{\it Spacetime Structure near Generic Horizons and Soft Hair}, 
PRL {\bf 124} (2020) 041601 [arXiv:1908.09833 [hep-th]].

\bibitem{Brown:1986nw}
J.~D.~Brown and M.~Henneaux, 
{\it Central Charges in the Canonical Realization of Asymptotic Symmetries: An Example from Three-Dimensional Gravity}, 
Commun.\ Math.\ Phys.\ {\bf 104} (1986) 207.

%\bibitem{Ashtekar:1997ay}
%A.~Ashtekar, J.~Bi\v{c}\'{a}k and B.~G.~Schmidt, 
%{\it Asymptotic structure of symmetry reduced general relativity}, 
%Phys.\ Rev.\ D {\bf 55}, 669 (1997) [gr-qc/9608042].

\bibitem{Barnich:2007cs}
G.~Barnich and G.~Comp\`{e}re, 
{\it Classical central extension for asymptotic symmetries at null infinity in three spacetime dimensions}, 
Class.\ Quant.\ Grav.\ {\bf 24} (2007) F15; {\bf 24} (2007) 3139 
[gr-qc/0610130].

\bibitem{Barnich:2014kra}
G.~Barnich and B.~Oblak, 
{\it Notes on the BMS group in three dimensions: I. Induced representations}, 
JHEP {\bf 06} (2014) 129 
[arXiv:1403.5803 [hep-th]].

\bibitem{Barnich:2015uva}
G.~Barnich and B.~Oblak, 
{\it Notes on the BMS group in three dimensions: II. Coadjoint representation}, 
JHEP {\bf 03} (2015) 033 
[arXiv:1502.00010 [hep-th]].

\bibitem{Parsa:2019os}
A.~Farahmand Parsa, H.~R.~Safari and M.~M.~Sheikh-Jabbari, 
{\it On rigidity of 3d asymptotic symmetry algebras}, 
JHEP {\bf 03}, 143 (2019) [arXiv:1809.08209 [hep-th]].

\bibitem{Safari:2019os}
H.~R.~Safari and M.~M.~Sheikh-Jabbari, 
{\it BMS$_4$ algebra, its stability and deformations}, 
JHEP {\bf 04} (2019) 68 
[arXiv:1902.03260 [hep-th]].

\bibitem{Compere:2019ts}
G.~Comp\`{e}re, A.~Fiorucci and R.~Ruzziconi, 
{\it The $\Lambda$-BMS$_4$ group of dS$_4$ and new boundary conditions for AdS$_4$}, 
Class.\ Quant.\ Grav.\ {\bf 36} (2019) 195017; {\bf 38} (2021) 229501
[arXiv:1905.00971 [gr-qc]].

\bibitem{Figueroa:2022ns}
J.~Figueroa-O'Farrill, 
{\it Non-lorentzian spacetimes}, 
Differ.\ Geom.\ Appl.\ {\bf 82} (2022) 101894 
[arXiv:2204.13609 [math.DG]].

\bibitem{Bergshoeff:2023ar}
E.~Bergshoeff, J.~Figueroa-O'Farrill and J.~Gomis, 
{\it A non-lorentzian primer}, 
SciPost Phys.\ Lect.\ Notes {\bf 69} (2023) 1 
[arXiv:2206.12177 [hep-th]].

\bibitem{Bacry:1968ps}
H.~Bacry and J.~L\'{e}vy-Leblond, 
{\it Possible kinematics}, 
J.\ Math.\ Phys.\ {\bf 9} (1968) 1605.

\bibitem{Bacry:1986cy}
H. Bacry and J. Nuyts, 
{\it Classification of ten-dimensional kinematical groups with space isotropy}, 
J.\ Math.\ Phys.\ {\bf 27} (1986) 2455.

\bibitem{Andrzejewski:2018ks}
T.~Andrzejewski and J.~Figueroa-O'Farrill, 
{\it Kinematical lie algebras in 2 + 1 dimensions}, 
J.\ Math.\ Phys.\ {\bf 59} (2018) 061703 
[arXiv:1802.04048 [hep-th]].

%\bibitem{Figueroa:2018ky} 
%J.~Figueroa-O'Farrill, 
%{\it Kinematical Lie algebras via deformation theory}, 
%J.\ Math.\ Phys.\ {\bf 59}, 061701 (2018) [arXiv:1711.06111 [hep-th]].

%\bibitem{Figueroa:2018hy}
%J.~Figueroa-O'Farrill, 
%{\it Higher-dimensional kinematical Lie algebras via deformation theory}, 
%J.\ Math.\ Phys.\ {\bf 59}, 061702 (2018) [arXiv:1711.07363 [hep-th]].

%\bibitem{Duval:1991cs}
%C.~Duval, G.~W.~Gibbons and P.~A.~Horv\'{a}thy, 
%{\it Celestial mechanics, conformal structures, and gravitational waves},
%Phys.\ Rev.\ D {\bf 43}, 3907 (1991).

\bibitem{Duval:2014ce}
C.~Duval, G.~W.~Gibbons, P.~A.~Horv\'{a}thy and P.~M.~Zhang, 
{\it Carroll versus Newton and Galilei: two dual non-Einsteinian concepts of time},
Class.\ Quant.\ Grav.\ {\bf 31} (2014) 085016 
[arXiv:1402.0657 [gr-qc]].

\bibitem{Figueroa:2023ly}
J.~Figueroa-O'Farrill, 
{\it Lie algebraic Carroll/Galilei duality}, 
J.\ Math.\ Phys.\ {\bf 64} (2023) 013503 
[arXiv:2210.13924 [math.DG]].

%\bibitem{Henneaux:1979gs}
%M.~Henneaux, 
%{\it Geometry of Zero Signature Space-times}, 
%Bull.\ Soc.\ Math.\ Belg.\ {\bf 31}, 47 (1979).

%\bibitem{Henneaux:2021cs}
%M.~Henneaux and P.~Salgado-Rebolled\'{o}, 
%{\it Carroll contractions of Lorentz-invariant theories}, 
%JHEP {\bf 11}, 180 (2021) [arXiv:2109.06708 [hep-th]].

%\bibitem{Isham:1976sy}
%C.~J.~Isham, 
%{\it Some quantum field theory aspects of the superspace quantization of general relativity}, 
%Proc.\ Roy.\ Soc.\ Lond.\ A {\bf 351}, 209 (1976).

%\bibitem{Belinsky:1970oy}
%V.~A.~Belinsky, I.~M.~Khalatnikov and E.~M.~Lifshitz,
%{\it Oscillatory approach to a singular point in the relativistic cosmology}, 
%Adv.\ Phys.\ {\bf 19}, 525 (1970).

%\bibitem{Andersson:2005as}
%L.~Andersson, H.~van Elst, W.~C.~Lim and C.~Uggla,
%{\it Asymptotic silence of generic cosmological singularities}, 
%Phys.\ Rev.\ Lett.\ {\bf 94}, 051101 (2005) [gr-qc/0402051].

%\bibitem{Mielczarek:2013ay}
%J.~Mielczarek, 
%{\it Asymptotic silence in loop quantum cosmology}, 
%AIP Conf.\ Proc.\ {\bf 1514}, 81 (2013) [arXiv:1212.3527 [gr-qc]].

%\bibitem{Mielczarek:2017se}
%J.~Mielczarek and T.~Trze\'{s}niewski, 
%{\it Spectral dimension with deformed spacetime signature}, 
%Phys.\ Rev.\ D {\bf 96}, 024012 (2017) [arXiv:1612.03894 [hep-th]].

\bibitem{Bergshoeff:2017cy}
E.~Bergshoeff, J.~Gomis, B.~Rollier, J.~Rosseel and T.~ter Veldhuis, 
{\it Carroll versus Galilei gravity}, 
JHEP {\bf 03} (2017) 165 
[arXiv:1701.06156 [hep-th]].

\bibitem{Hansen:2020nr}
D.~Hansen, J.~Hartong and N.~A.~Obers, 
{\it Non-relativistic gravity and its coupling to matter}, 
JHEP {\bf 06} (2020) 145 
[arXiv:2001.10277 [gr-qc]].

\bibitem{Hansen:2022cy}
D.~Hansen, N.~A.~Obers, G.~Oling and B.~T.~S\o gaard, 
{\it Carroll expansion of general relativity}, 
SciPost Phys.\ {\bf 13} (2022) 055 
[arXiv:2112.12684 [hep-th]].

\bibitem{Hartong:2023ry}
J.~Hartong, N.~A.~Obers and G.~Oling, 
{\it Review on non-relativistic gravity}, 
Front.\ in Phys.\ {\bf 11} (2023) 1116888 
[arXiv:2212.11309 [gr-qc]].

\bibitem{Bleeken:2017ty}
J.~Van den Bleeken, 
{\it Torsional Newton-Cartan gravity from the large c expansion of general relativity}, 
Class.\ Quant.\ Grav.\ {\bf 34} (2017) 185004 
[arXiv:1703.03459 [gr-qc]].

\bibitem{Hansen:2019ay}
D.~Hansen, J.~Hartong and N.~A.~Obers, 
{\it Action principle for Newtonian gravity}, 
Phys.\ Rev.\ Lett.\ {\bf 122} (2019) 061106 
[arXiv:1807.04765 [hep-th]].

\bibitem{Papageorgiou:2009as}
G.~Papageorgiou and B.~J.~Schroers, 
{\it A Chern-Simons approach to Galilean quantum gravity in 2+1 dimensions}, 
JHEP {\bf 11} (2009) 009 
[arXiv:0907.2880 [hep-th]].

\bibitem{Papageorgiou:2010ga}
G.~Papageorgiou and B.~J.~Schroers, 
{\it Galilean quantum gravity with cosmological constant and the extended $q$-Heisenberg algebra}, 
JHEP {\bf 11} (2010) 020 
[arXiv:1008.0279 [hep-th]].

\bibitem{Matulich:2019ln}
J.~Matulich, S.~Prohazka and J.~Salzer, 
{\it Limits of three-dimensional gravity and metric kinematical Lie algebras in any dimension}, 
JHEP {\bf 07} (2019) 118 
[arXiv:1903.09165 [hep-th]].

\bibitem{Gomis:2020ny}
J.~Gomis, A.~Kleinschmidt, J.~Palmkvist and P.~Salgado-Rebolledo, 
{\it Newton-Hooke/Carrollian expansions of (A)dS and Chern-Simons gravity}, 
JHEP {\bf 02} (2020) 009 
[arXiv:1912.07564 [hep-th]].

\bibitem{Duval:1985by}
C.~Duval, G.~Burdet, H.~P.~K\"{u}nzle and M.~Perrin, 
{\it Bargmann structures and Newton-Cartan theory}, 
Phys.\ Rev.\ D {\bf 31} (1985) 1841.

\bibitem{Gibbons:2003nt}
G.~W.~Gibbons and C.~E.~ Patricot, 
{\it Newton-Hooke spacetimes, Hpp-waves and the cosmological constant}, 
Class.\ Quant.\ Grav.\ {\bf 20} (2003) 5225 
[hep-th/0308200].

\bibitem{Figueroa:2019ss}
J.~Figueroa-O'Farrill and S.~Prohazka, 
{\it Spatially isotropic homogeneous spacetimes}, 
JHEP {\bf 01} (2019) 229 
[arXiv:1809.01224 [hep-th]].

%\bibitem{Duval:2017cs}
%C.~Duval, G.~W.~Gibbons, P.~A.~Horv\'{a}thy and P.~M.~Zhang, 
%{\it Carroll symmetry of plane gravitational waves}, 
%Class.\ Quant.\ Grav.\ {\bf 34}, 175003 (2017) [arXiv:1702.08284 [gr-qc]].

%\bibitem{Morand:2020es}
%K.~Morand, 
%{\it Embedding Galilean and Carrollian geometries. I. Gravitational waves}, 
%J.\ Math.\ Phys.\ {\bf 61}, 082502 (2020) [arXiv:1811.12681 [hep-th]].

%\bibitem{Donnay:2019cn}
%L.\ Donnay and C.\ Marteau, 
%{\it Carrollian physics at the black hole horizon}, 
%Class.\ Quant.\ Grav.\ {\bf 36}, 165002 (2019) [arXiv:1903.09654 [hep-th]].

\bibitem{Donnay:2023by}
L.~Donnay, A.~Fiorucci, Y.~Herfray and R.~Ruzziconi, 
{\it Bridging Carrollian and celestial holography}, 
Phys.\ Rev.\ D {\bf 107} (2023) 126027 
[arXiv:2212.12553 [hep-th]].

\bibitem{Taylor:2016ly}
M.~Taylor, 
{\it Lifshitz holography}, 
Class.\ Quant.\ Grav.\ {\bf 33} (2016) 033001 
[arXiv:1512.03554 [hep-th]].

\bibitem{Perez:2021ay}
A.~P\'{e}rez, 
{\it Asymptotic symmetries in Carrollian theories of gravity}, 
JHEP {\bf 12} (2021) 173 
[arXiv:2110.15834 [hep-th]].

\bibitem{Fuentealba:2022as}
O.~Fuentealba, M.~Henneaux, P.~Salgado-Rebolledo and J.~Salzer, 
{\it Asymptotic structure of Carrollian limits of Einstein-Yang-Mills theory in four spacetime dimensions}, 
Phys.\ Rev.\ D {\bf 106} (2022) 104047 
[arXiv:2207.11359 [hep-th]].

\bibitem{Freidel:2023bnj}
L.~Freidel, M.~Geiller and W.~Wieland, 
{\it Corner symmetry and quantum geometry}, 
arXiv:2302.12799 [hep-th].

\bibitem{Ciambelli:2023mir}
L.~Ciambelli, L.~Freidel and R.~G.~Leigh, 
{\it Null Raychaudhuri: Canonical Structure and the Dressing Time}, 
arXiv:2309.03932 [hep-th].

\bibitem{Adami:2023ce}
H.~Adami, A.~Parvizi, M.~M.~Sheikh-Jabbari, V.~Taghiloo and H.~Yavartanoo, 
{\it Carrollian Structure of the Null Boundary Solution Space}, 
arXiv:2311.03515 [hep-th].

\bibitem{Duval:2014cs}
C.~Duval, G.~W.~Gibbons and P.~A.~Horv\'{a}thy, 
{\it Conformal Carroll groups and BMS}, 
Class.\ Quant.\ Grav.\ {\bf 31} (2014) 092001 
[arXiv:1402.5894 [gr-qc]].

\bibitem{Ciambelli:2019cs}
L.~Ciambelli, R.~G.~Leigh, C.~Marteau, and P.~M.~Petropoulos, 
{\it Carroll structures, null geometry, and conformal isometries}, 
Phys.\ Rev.\ D {\bf 100} (2019) 046010 
[arXiv:1905.02221 [hep-th]].

\bibitem{Matschull:1997du}
H.~J.~Matschull and M.~Welling,
{\it Quantum mechanics of a point particle in (2+1)-dimensional gravity}, 
Class.\ Quant.\ Grav.\ {\bf 15} (1998) 2981 
[arXiv:gr-qc/9708054 [gr-qc]].

\bibitem{Kowalski:2014dy}
J.~Kowalski-Glikman and T.~Trze\'{s}niewski, 
{\it Deformed Carroll particle from 2+1 gravity}, 
Phys.\ Lett.\ B {\bf 737}, 267 (2014) 
[arXiv:1408.0154 [hep-th]].

\bibitem{Trzesniewski:2018ey}
T.~Trze\'{s}niewski, 
{\it Effective Chern-Simons actions of particles coupled to 3D gravity}, 
Nucl.\ Phys.\ B {\bf 928}, 448 (2018) 
[arXiv:1706.01375 [hep-th]].

\bibitem{Trzesniewski:2023gs}
T.~Trze\'{s}niewski, 
{\it 3D Gravity, Point Particles, and Deformed Symmetries}, 
Acta Phys.\ Polon.\ B Proc.\ Suppl.\ {\bf 16}, 6-A19 (2023) 
[arXiv:2212.14031 [hep-th]].

\bibitem{Borowiec:2017ag}
A.~Borowiec, J.~Lukierski and V.~N.~Tolstoy, 
{\it Addendum to ``Quantum deformations of $D = 4$ Euclidean, Lorentz, Kleinian and quaternionic $\mathfrak{o}^\star(4)$ symmetries in unified $\mathfrak{o}(4;\mathbbm{C})$ setting''}, 
Phys.\ Lett.\ B {\bf 770} (2017) 426 
[arXiv:1704.06852 [hep-th]].

\bibitem{Kowalski:2020qs}
J.~Kowalski-Glikman, J.~Lukierski and T.~Trze\'{s}niewski, 
{\it Quantum D = 3 Euclidean and Poincar\'{e} symmetries from contraction limits}, 
JHEP {\bf 09} (2020) 096 
[arXiv:1911.09538 [hep-th]].

\bibitem{Perez:2022at}
A.~P\'{e}rez, 
{\it Asymptotic symmetries in Carrollian theories of gravity with a negative cosmological constant}, 
JHEP {\bf 09} (2022) 044 
[arXiv:2202.08768 [hep-th]].

%\bibitem{Bengtsson:1983cn}
%A.~K.~H.~Bengtsson, I.~Bengtsson and L.~Brink,
%{\it Cubic interaction terms for arbitrary spin},
%Nucl.\ Phys.\ B {\bf 227} (1983) 31.

\bibitem{Borowiec:2019ky}
A.~Borowiec, L.~Brocki, J.~Kowalski-Glikman and J.~Unger, 
{\it $\kappa$-deformed BMS symmetry}, 
Phys.\ Lett.\ B {\bf 790} (2019) 415 
[arXiv:1811.05360 [hep-th]].

\bibitem{Borowiec:2021bs}
A.~Borowiec, L.~Brocki, J.~Kowalski-Glikman and J.~Unger, 
{\it BMS algebras in 4 and 3 dimensions, their quantum deformations and duals}, 
JHEP {\bf 02} (2021) 084 
[arXiv:2010.10224 [hep-th]].

\bibitem{Borowiec:2021ds}
A.~Borowiec, J.~Kowalski-Glikman and J.~Unger, 
{\it 3-dimensional $\Lambda$-BMS symmetry and its deformations}, 
JHEP {\bf 11} (2021) 103 
[arXiv:2106.12874 [hep-th]].

\bibitem{Trzesniewski:2023qi}
T.~Trze\'{s}niewski, 
{\it Quantum symmetries in 2+1 dimensions: Carroll, (a)dS-Carroll, Galilei and (a)dS-Galilei}, 
JHEP {\bf 02} (2024) 200 
[arXiv:2306.05409 [hep-th]].

\bibitem{Ballesteros:2020ts}
A.~Ballesteros, G.~Gubitosi, I.~Gutierrez-Sagredo and F.~J.~Herranz, 
{\it The $\kappa$-Newtonian and $\kappa$-Carrollian algebras and their noncommutative spacetimes}, 
Phys.\ Lett.\ B {\bf 805} (2020) 135461 
[arXiv:2003.03921 [hep-th]].

%\bibitem{Bergshoeff:2014ds}
%E.~Bergshoeff, J.~Gomis and G.~Longhi, 
%{\it Dynamics of Carroll particles},
%Class.\ Quant.\ Grav.\ {\bf 31}, 205009 (2014) [arXiv:1405.2264 [hep-th]].

%\bibitem{Cardona:2016ds}
%B.~Cardona, J.~Gomis and J.~M.~Pons, 
%{\it Dynamics of Carroll strings},
%JHEP {\bf 07}, 50 (2016) [arXiv:1605.05483 [hep-th]].

%\bibitem{Marsot:2022pn}
%L.~Marsot, 
%{\it Planar Carrollean dynamics, and the Carroll quantum equation},
%J.\ Geom.\ Phys.\ {\bf 179}, 104574 (2022) [arXiv:2110.08489 [math-ph]].

%\bibitem{Marsot:2023hs}
%L.~Marsot, P.~M.~Zhang, M.~Chernodub and P.~A.~Horv\'{a}thy, 
%{\it Hall motions in Carroll dynamics},
%Phys.\ Rept.\ {\bf 1028}, 1 (2023) [arXiv:2212.02360 [hep-th]].

\end{thebibliography}
\end{document}